\documentclass{aastex}
\usepackage{emulateapj5}    
\usepackage{graphicx}
\newcommand{\be}{\begin{equation}}
\newcommand{\ee}{\end{equation}}
\newcommand{\bea}{\begin{eqnarray}}
\newcommand{\eea}{\end{eqnarray}}

\begin{document}
\title{Simulations of electron Magnetohydrodynamic Turbulence}
\shorttitle{EMHD Turbulence}
\shortauthors{Cho \& Lazarian}

\author{Jungyeon Cho\altaffilmark{1,2} 
\and
A. Lazarian\altaffilmark{2}
}
\altaffiltext{1}{Dept. of Astronomy and Space Science, 
    Chungnam National Univ., Daejeon, Korea; cho@canopus.cnu.ac.kr}
\altaffiltext{2}{Dept. of Astronomy, Univ. of Wisconsin, Madison, 
    WI53706, USA; lazarian@astro.wisc.edu}

\begin{abstract}
We present numerical simulations of electron magnetohydrodynamic (EMHD)
and electron reduced MHD (ERMHD) turbulence.
Comparing scaling relations,
we find that both EMHD and ERMHD turbulence show
similar spectra and anisotropy.
We develop new techniques to study anisotropy of EMHD turbulence.
Our detailed study of anisotropy of EMHD turbulence supports
our earlier result of $k_{\|}\propto k_{\perp}^{1/3}$ scaling, where
$k_{\|}$ and $k_{\perp}$ are wavenumbers parallel and perpendicular 
to local direction of magnetic field, respectively. 
We find that the
 high-order statistics show a scaling that is similar to the She-Leveque scaling for 
incompressible hydrodynamic turbulence and different from that of
incompressible MHD turbulence. We observe that the bispectra, which characterize
the interaction of different scales within the turbulence cascade, are very
different for EMHD and MHD turbulence. 
We show that both decaying and driven EMHD turbulence have the same statistical
properties.
We calculate the probability distribution functions (PDFs) 
of MHD and EMHD turbulence and compare them with
those of interplanetary turbulence.
We find that, as in the case of the solar wind, 
the PDFs of the increments of magnetic field strength in MHD and EMHD
turbulence
are well described by the Tsallis distribution.
We discuss implications of our results for astrophysical situations, including
the advection dominated accretion flows and magnetic reconnection.
\end{abstract}

\section{Introduction}
Turbulence at scales below the proton gyroradius is of great importance
in many astrophysical applications. Such turbulence involving motions of
electrons is essential for understanding the small-scale magnetic field dynamics of plasmas. 
It is
also important for understanding of magnetic fields in the crust of a neutron star 
(Goldreich \& Reisenegger 1992). This turbulence has been measured at by solar wind probes 
(Leamon et al. 1998).

The origin of the small-scale turbulence in a magnetized plasma
is easy to understand if we think of what is happening
with turbulent motions at small scales.
When turbulence is driven on large scales, turbulence energy cascades
down to smaller scales.
The nature of magnetized turbulence from the outer scale to the proton
gyroradius scale is relatively well known.
Magnetized turbulence above the proton gyroradius
can be described by the standard magnetohydrodynamics (MHD).
As the name implies, the standard MHD treats the plasma as a single fluid.
MHD turbulence can be
decomposed into cascades of Alfven, fast and slow modes 
(Goldreich \& Sridhar 1995; Lithwick \& Goldreich 2001; Cho \& Lazarian 2002, 2003).
While
fast and slow modes get damped at larger scales, Alfvenic modes
can cascade down to the proton gyroradius scale. 
Near and below the proton gyroradius scale,
single-fluid description fails and we should take into account kinetic effects.
Then what will happen to the Alfven modes that reach the proton gyroradius scale?

Recent years, the nature of small-scale MHD turbulence in the solar wind has drawn 
 interest from researchers (Howes et al. 2008a,b; 
Matthaeus, Servidio, \& Dmitruk 2008; Saito et al. 2008; Gary, Saito, \& Li 2008;
Schekochihin et al. 2009; Dmitruk \& Matthaeus 2006).
In situ measurements of the solar wind show magnetic fluctuations over a broad range of
frequencies. In the rest frame of the spacecrafts, the magnetic fluctuations show
a broken power-law spectrum. For example, Leamon et al. (1999) reported that,
at $\sim 0.2Hz$, the spectrum breaks from a $\nu^{-1.67}$ power-law to
a $\nu^{-2.91}$ power-law.
In general, the spectral break-point lies in the range $0.2 Hz \lesssim \nu \lesssim 0.5 Hz$
(see Saito et al. 2008).
The $\nu^{-1.67}$ power-law at $\nu \lesssim 0.2Hz$ seems to be relatively robust and
represents inertial range of Alfvenic MHD turbulence.
However, the power index for $\nu \gtrsim 0.5Hz$ seems to vary between
-2 and -4 (Leamon et al. 1998; Smith et al. 2006).
This range, characterized by a steeper power-law index, is termed ``dispersion range"
(Stawicki, Gary, \& Li 2001), which is different from the dissipation range.
The true dissipation scale of turbulence may lie at the end of the dispersion range.
When we convert frequency to length scale, the broken power-law implies that
the magnetic energy spectrum changes from a $k^{-1.67}$ inertial range spectrum
to a steeper dispersion range spectrum,
as we move from large scales to small scales.
The transition from the inertial range to the dispersion range occurs near the proton gyro-scale
$\rho_i$.
(However, it is also possible that it occurs at the ion inertial scale 
$d_i=\rho_i/\sqrt{\beta_i}$, where $\beta_i$ is the ion plasma $\beta$. See discussion in
Schekochihin et al. 2009).
The identity of the dispersion range turbulence is still under debate.

Small-scale magnetized turbulence also plays 
important roles in other astrophysical objects.
Among them, the crust of neutron stars gives us useful insights
on the electron MHD (EMHD) model of small-scale magnetized turbulence.
In the crust of neutron stars, ions are virtually immobile and, thus,
the ion gyroradius scale can be regarded as infinite.
Therefore, turbulence in the crust should be similar to small-scale turbulence.
Since ions are immobile, they provide a smooth charge background and electrons carry 
all the current, so that
\be
   {\bf v}_e = -\frac{ {\bf J} }{ n_e e }=-\frac{c}{4\pi n_e e} \nabla \times {\bf B},
  \label{eq1}
\ee
where ${\bf v}_e$ is the electron velocity,
${\bf J}$ is electric current density,
${\bf B}$ is magnetic field, $c$ is the speed of light,
$n_e$ is the electron number density, and $e$ is the absolute value of the electric charge.
Inserting this relation into the magnetic induction equation, we can obtain the
EMHD equation
\begin{equation}
   \frac{ \partial {\bf B} }{ \partial t }
  = - \frac{ c }{ 4 \pi n_e e } \nabla \times \left[
    (\nabla \times {\bf B}) \times{\bf B} \right] + \eta \nabla^{2} {\bf B},
\end{equation}
where $\eta$ is magnetic diffusivity (see Kingsep, Chukbar, \& Yankov 1990 for
details about EMHD).
Goldreich \& Reisenegger (1992) first showed that magnetized turbulence 
in the crust of neutron stars can be
described by the EMHD equation.
They discussed
the properties of EMHD turbulence in neutron stars and argued 
that EMHD turbulence can enhance
ohmic dissipation of magnetic field in isolated neutron stars (see also
Cumming, Arras, \& Zweibel 2004).

In this paper, we will focus on spectrum and anisotropy of EMHD turbulence. 
Earlier researchers convincingly showed that
energy spectrum of EMHD turbulence is steeper than 
Kolmogorov's $k^{-5/3}$ spectrum (Biskamp, Schwarz, \& Drake 1996;
Biskamp et al. 1999; Ng et al.2003).
They found that 
the energy spectrum follows 
\begin{equation}
  E(k) \propto k^{-7/3}.
\end{equation}
The steep energy spectrum can be explained by
 the following Kolmogorov-type
argument (Biskamp et al. 1996).
Suppose that the eddy interaction time for eddies of size $l$ is 
the usual eddy turnover time
$t_{cas,l} \sim l/v_l$.
Since ${\bf v}\propto \nabla \times {\bf B}$ (Eq.~[\ref{eq1}]),
this becomes $t_{cas,l} \propto l^2/b_l$.
Combining this with the constancy of
spectral energy cascade rate ($b_l^2/t_{cas,l}$=constant),
one obtains 
$E(k)\propto k^{-7/3}$. 
Note that $E(k)$ and $b_l$ are related by $kE(k) \sim b_l^2$.
Since earlier researchers convincingly obtained the $k^{-7/3}$ spectrum,
more focus is given to anisotropy of EMHD turbulence.

Cho \& Lazarian (2004; hereinafter CL04) derived the expression for anisotropy
\begin{equation}
k_{\|}\propto k_{\bot}^{1/3}
\label{ani}
\end{equation}
where $k_{\bot}$ and $k_{\|}$ should be understood as wavenumbers parallel and perpendicular
to the {\it local} magnetic field, the same way as parallel and perpendicular wavenumbers are
understood in Goldreich-Sridhar (1995) model of MHD turbulence\footnote{A wavelet description
would be more appropriate, as usual wavenumbers are defined in the {\it global} magnetic field frame. 
We keep this in mind, while using the traditional wavenumber notations.}. 
The expression (\ref{ani}), as we discuss in \S3, follows from the application of the critical balance 
notion to the electron MHD cascade.  

In this paper, we only consider strong EMHD turbulence.
Discussions on weak EMHD turbulence can be found in
 Galtier \& Bhattacharjee (2003) and Galtier (2006).
In \S2, we compare EMHD and recently proposed ERMHD (Schekochihin et al. 2009) formalisms. 
In \S3, we discuss expected scaling relations of EMHD and ERMHD turbulence.
In \S4, we describe our numerical methods.
In \S5, we compare EMHD and ERMHD turbulence.
In the section we perform simulations in an elongated numerical box ($786\times 256^2$).
In \S6, we present detailed study of anisotropy
{}from an EMHD simulation with $512^3$ resolution.
In \S7, we present high-order statistics and bispectra of EMHD turbulence.
In \S8, we calculate the probability distribution functions (PDFs) and
        briefly compare them with the solar wind data.
We give discussions in \S9 and summary in \S10.

\section{EMHD and ERMHD}
Alfven modes are incompressible and thus less prone to collisionless damping 
(Barnes 1966; Kulsrud \& Pearse 1969).
Therefore, in the solar wind, 
it is generally accepted that Alfvenic turbulence provides
a suitable description of fluid motions on scales larger 
than the proton gyroscale.
The physics of strong Alfven turbulence is relatively well understood 
(see, e.g., Goldreich \& Sridhar 1995)

However, turbulence on scales smaller than the proton gyroscale
is not well understood.
Nevertheless, there are several models for this small-scale turbulence.
Here, we consider only fluid-like models of the small-scale turbulence.
In particular, we consider Electron MHD (EMHD) model and 
Electron Reduced MHD (ERMHD) model.
The former has been studied since 1990s (Kingsep et al. 1990; Biskamp et al. 1996) 
and the latter has been proposed
only recently (Schekochihin et al. 2009).
Both models can be derived from the generalized Ohm's law.

\subsection{EMHD}
EMHD can be viewed Hall MHD in the limit of $k\rho_i \gg 1$, where
$\rho_i$ is the ion gyroradius.
Hall MHD equations are very similar to the standard MHD ones. 
But, there are also differences.
The most important difference is that Hall MHD is based on
the generalized Ohm's law
\be
   {\bf E}=-\frac{{\bf v}}{c} \times {\bf B} +\frac{ {\bf J}\times {\bf B} }{n_eec}
  +  \frac{\bf J}{\sigma},
\ee
while the standard MHD uses the `standard' Ohm's law
\be
   {\bf E}=-\frac{{\bf v}}{c} \times {\bf B} + \frac{\bf J}{\sigma}.
\ee
Here, ${\bf v}$ is fluid velocity, 
${\bf E}$ is electric field,
and $\sigma$ is
conductivity.
Compared with that of the standard MHD,
the induction equation of Hall MHD has an extra term (the ${\bf J}\times {\bf B}$ term on 
the right-hand side):
\bea
 \frac{ \partial {\bf B} }{ \partial t }&=& -c\nabla \times {\bf E}, \nonumber \\
                                        &=&\nabla \times ({\bf v}\times {\bf B})
   - \nabla \times \frac{ {\bf J}\times {\bf B} }{n_ee} +\frac{c^2}{4\pi\sigma} \nabla^2{\bf B} \nonumber \\
                                        &=&\nabla \times ({\bf v}\times {\bf B})
   - \nabla \times \frac{ c(\nabla \times {\bf B})\times {\bf B} }{4\pi n_ee} +\eta \nabla^2{\bf B}, \mbox{~~~}
  \label{eq:inductionh}  
\eea
where $\eta=c^2/(4\pi\sigma)$ and we use $\nabla \times {\bf B}=(4\pi/c){\bf J}$.

When we normalize magnetic field to a velocity (i.e.~$
{\bf B}/\sqrt{4\pi\rho}={\bf B}/\sqrt{4\pi n_i m_i}\rightarrow {\bf B}$, 
where $n_i$ is the ion number density and
$m_i$ is the ion mass), 
Eq.~(\ref{eq:inductionh}) becomes
\be
  \frac{ \partial {\bf B} }{ \partial t }
     =\nabla \times ({\bf v}\times {\bf B})
    - d_i \nabla \times \left[ (\nabla \times {\bf B})\times {\bf B} \right]  
    +\eta \nabla^2{\bf B}, \mbox{~~(Hall MHD)}
  \label{eq:norm_hmhd}
\ee
where
\be
   d_i = \frac{ c\sqrt{4\pi n_i m_i} }{4\pi n_e e}=\frac{c}{\sqrt{4\pi n_i (Ze)^2/m_i}}
        =\frac{c}{\omega_{pi}}=\frac{\rho_i}{\sqrt{\beta_i}}.
    \label{eq:9}
\ee
Here, $\rho_i$ is the ion gyroradius, $Z=n_e/n_i=q_i/e$, and $\beta_i$ is the ion plasma beta:
\be
  \beta_i = \frac{ n_ik_BT }{ B^2/8\pi },
\ee
where $k_B$ is the Boltzmann constant.
The order of magnitude values of the first and the second terms on the right-hand side are
\bea
    \nabla \times ({\bf v}\times {\bf B}) \sim b^2/l, \\
    d_i \nabla \times \left[ (\nabla \times {\bf B})\times {\bf B} \right] \sim d_i b^2/l^2,
\eea 
where $l$ is the scale of interest and we assume $v\sim b$.
Eq.~(\ref{eq:norm_hmhd}) reduces to the standard MHD induction equation for $l\gg d_i$,
 while it reduces to the Electron MHD equation for $l\ll d_i$:
\be
  \frac{ \partial {\bf B} }{ \partial t }
     =
    - d_i \nabla \times \left[ (\nabla \times {\bf B})\times {\bf B} \right]  +\eta \nabla^2{\bf B}. \mbox{~~(EMHD)}
  \label{eq:emhd}
\ee

In usual collisionless plasmas, 
transition from standard MHD to EMHD occurs at the ion inertial scale
$d_i$.
When ions are immobile and provide a homogeneous background, 
as in the crust of a neutron star, 
EMHD can be applicable even at the outer scale of turbulence, which 
can be larger than $d_i$ defined by the first two expressions in Eq.~(\ref{eq:9}).
Note that
anisotropy of turbulence (i.e. $k_{\perp} \gg k_{\|}$) at $d_i$ (in usual plasmas) or 
on the energy injection scale (in case ions are immobile)
is not a necessary condition for EMHD.

\subsection{ERMHD}
Schekochihin et al.~(2009) first derived ERMHD from kinetic RMHD equations.
They also gave derivation of ERMHD equations
{}from the generalized Ohm's law.
Therefore the starting point of ERMHD may be also the generalized Ohm's law and
derivation of ERMHD 
is identical to the EMHD case
 up to Eq.~(\ref{eq:inductionh}) (see previous subsection).
Note that
the velocity ${\bf v}$ in Eq.~(\ref{eq:inductionh}) 
denotes ion velocity ${\bf v}_i$ and that
RMHD, hence ERMHD, assumes anisotropy of turbulence
(i.e.~ $k_{\perp} \gg k_{\|}$).

However, ERMHD assumes that the term
\be
   \nabla \times ({\bf v}\times {\bf B}) \approx -{\bf B}\nabla \cdot {\bf v}
   \label{eq:divv}
\ee
in
Eq.~(\ref{eq:inductionh}) may not be negligible 
in the limit of $k\rho_i \gg 1$.
That is, ERMHD assumes ${\bf v}_i \approx 0$ in this limit, but 
$\nabla \cdot {\bf v}_i=\nabla \cdot {\bf v}_e \ne 0$
(see Schekochihin et al.~2009).
Taking this into account, one can rewrite the magnetic induction equation as
\be
   \frac{\partial {\bf B}}{\partial t}
    = -{\bf B}\nabla \cdot {\bf v}_i 
          -\frac{c}{4\pi e n_{e}}\nabla \times[{\bf B}\cdot \nabla {\bf B}], 
\ee
which becomes
\bea
   \frac{ \partial {\bf B}_{\perp} }{\partial t} =
      -\frac{c}{4\pi e n_{e}}\nabla_{\perp} \times[{\bf B}\cdot \nabla {\bf B}_{\|}], 
  \label{eq:perp_3} \\
   \frac{ \partial {\bf B}_{\|} }{\partial t} =
      -{\bf B}_0 \nabla \cdot {\bf v}_i
      -\frac{c}{4\pi e n_{e}}\nabla_{\perp} \times[{\bf B}\cdot \nabla {\bf B}_{\perp}],
  \label{eq:par_3}  
\eea
where ${\bf B}_{\|}$ and ${\bf B}_{\perp}$ denote the components of magnetic field
parallel and perpendicular to the mean field ${\bf B}_0$, respectively
(see Appendix A of this paper and Appendix C of Schekochihin et al.~2009).
Here we drop the dissipation term for simplicity.

Finally, from the continuity equation and the assumption of pressure balance, 
we can rewrite the above equations as
\bea
   \frac{ \partial {\bf B}_{\perp} }{\partial t} =
     -\frac{c}{4\pi en_e}\nabla_{\perp} \times[{\bf B}\cdot \nabla {\bf B}_{\|}],\\
   \frac{ \partial {\bf B}_{\|} }{\partial t} =
      -\frac{\beta_i(1+Z/\tau)}{2+\beta_i(1+Z/\tau)}
      \frac{c}{4\pi e n_e}\nabla_{\perp} \times[{\bf B}\cdot \nabla {\bf B}_{\perp}]
\eea
(see Eqs.~[\ref{eq:perp3}] and [\ref{eq:par4}]), which become
\be
  \frac{\partial}{\partial t} \frac{\tilde{\bf b}}{\sqrt{4\pi n_im_i}}
  =\frac{-\rho_i}{\sqrt{\beta_i}}
   \sqrt{\alpha}
   \nabla_{\perp} \times (\frac{{\bf B}}{\sqrt{4\pi n_im_i}}
     \cdot \nabla \frac{\tilde{\bf b}}{\sqrt{4\pi n_i m_i}}), \label{eq:ermhd}
\ee
where
\bea
   {\bf B}={\bf B}_0 + {\bf b}, \\
   {\bf b}={\bf b}_{\perp} + b_{\|}\hat{\bf z}, \\
   \tilde{\bf b}
  ={\bf b}_{\perp}+\sqrt{\frac{1}{\alpha}} b_{\|}\hat{\bf z}, \label{eq_tildeb} \\
   \alpha \equiv \frac{\beta_i (1+Z/\tau)}{2+\beta_i(1+Z/\tau)}. \label{eq_alpha}
\eea
Here, ${\bf B}_0$ ($=B_0 \hat{\bf z}$) is the mean magnetic field,
$\hat{\bf z}$ is a unit vector along the direction of ${\bf B}_0$,
$Z=q_i/e$ ($e=|q_e|$) is the ion-to-electron charge ratio,
$\tau=T_i/T_e$ is the temperature ratio, $n_e$ is the average electron number density.
The vector ${\bf b}_{\perp}$ (in Fourier space) 
is parallel to $\hat{\bf z} \times \hat{\bf k}_{\perp}$,
where $\hat{\bf k}_{\perp}={\bf k}_{\perp}/k_{\perp}$.
Therefore, in Fourier space $\tilde{\bf b}_{\bf k}$ lies in the plane spanned by 
$\hat{\bf z} \times \hat{\bf k}_{\perp}$ and $\hat{\bf z}$, which means
$\tilde{\bf b}_{\bf k}$ is perpendicular to ${\bf k}_{\perp}$.

\section{Expected Scaling Relations}

In the presence of a strong mean magnetic field, 
turbulence energy tends to cascade in
the direction perpendicular to the mean field.
As a result, in Fourier space, modes with $k_{\perp} \gg k_{\|}$ are predominantly
excited.
In real space, characteristic scales parallel to the mean field ($l_{\|}$) tend to be
larger than those perpendicular to it ($l_{\perp}$).
This is referred to as anisotropy of turbulence.

In the case of {\it standard} MHD turbulence, this global anisotropy
has been studied since early 1980s (Shebalin, Matthaeus, \& Montgomery 1983).
On the other hand, Goldreich \& Sridhar (1995) showed that
there exists a regime of turbulence in which
a critical balance is maintained between wave motions (with timescale of $t_w\sim l_{\|}/V_A$)
and hydrodynamic motions (with timescale of $l_{\perp}/v_l$).
This is the so-called strong turbulence regime and they found
a certain relation between $k_{\|}$ and $k_{\perp}$, $k_{\|} \propto k_{\perp}^{2/3}$,
in the regime.
This scale-dependent anisotropy was numerically confirmed by
Cho \& Vishniac (2000) and Maron \& Goldreich (2001).
Cho \& Vishniac (2000) showed that this scale-dependent anisotropy
can be measured only in a
{\it local} coordinate frame which is aligned with the {\it locally averaged}
magnetic field direction.  
The necessity of using a local frame is due to the fact that
eddies are aligned along the local mean magnetic field, rather than
the global mean field ${\bf B}_0$. 
We call this kind of anisotropy as {\it local} anisotropy. 

In the case of {\it EMHD} turbulence, anisotropy has been studied only recently.
Dastgeer et al.~(2000) and 
Dastgeer \& Zank (2003) numerically
studied global anisotropy of 2D EMHD turbulence.
On the other hand, Ng et al.~(2003) numerically studied local anisotropy
of 2D EMHD turbulence, but used second-order structure functions, the limitation of
which will be discussed in Section \S\ref{sect:sf2}.
In CL04, we studied 
anisotropy of 3D EMHD turbulence.
In CL04, we for the first time found that
critical balance between wave motions and hydrodynamic motions also holds true 
in EMHD turbulence and that anisotropy of EMHD,
$k_{\|} \propto k_{\perp}^{1/3}$, is stronger than that of standard MHD turbulence.

Anisotropy of {\it ERMHD} is expected to be
similar to that of EMHD (Schekochihin et al. 2009).
This is understandable from Eq.~(\ref{eq:ermhd}).
Note that 
${\bf b}={\bf b}_{\perp}+b_{\|} \hat{\bf z}$ and
$\tilde{\bf b}={\bf b}_{\perp}+\sqrt{1/\alpha}b_{\|} \hat{\bf z}$
in the equation. The factor $\alpha$ is less than or equal to $1$.
If we assume ${b}_{\perp} \sim \sqrt{1/\alpha}b_{\|}$, 
then the parallel (or $z$) component of true magnetic field $b_{\|}$ 
is equal to $\sim \sqrt{\alpha} b_{\perp} \le b_{\perp}$.
If $\alpha \sim 1$ and, hence, ${\bf b} \sim \tilde{\bf b}$, 
then Eq.~(\ref{eq:ermhd})
becomes the usual EMHD equation. Therefore it is trivial to show that
anisotropy of ERMHD is similar to that of EMHD.
On the other hand, if $\alpha \ll 1$ and, hence, 
${\bf b} ={\bf b}_{\perp}+\sqrt{\alpha}\tilde{\bf b}_{\|} \sim {\bf b}_{\perp}$,
then the ERMHD equation becomes different from the EMHD equation.
However, in general, the parallel component of ${\bf b}$ does not contribute much to
the term ${\bf B}\cdot \nabla \tilde{\bf b}$, because
$b_{\|} \hat{\bf z}\cdot \nabla \sim b_{\|} k_{\|}  
   \ll {\bf b}_{\perp}\cdot \nabla \sim b_{\perp}k_{\perp} $ in the presence
of anisotropy, $k_{\|} \ll k_{\perp}$.
This means that what matters for energy cascade is the perpendicular component
of ${\bf b}$, which is not directly affected by the value of $\alpha$.
Therefore, even in case of $\alpha \ll 1$, 
we expect that anisotropy of ERMHD is similar to that of EMHD.
However, this conjecture needs to be tested by numerical calculations.

\begin{figure*}[h!t]
\includegraphics[width=0.48\textwidth]{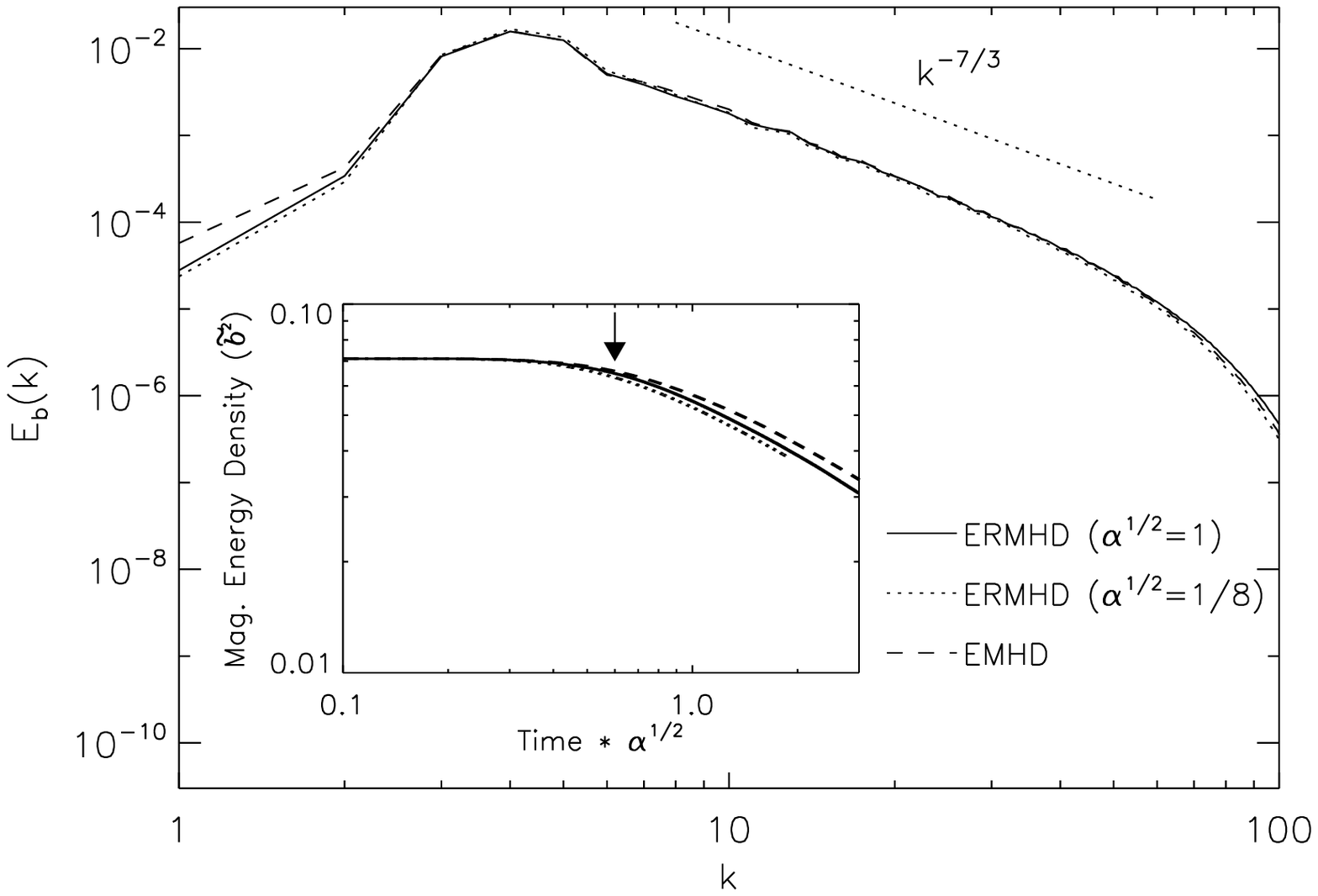}  
\includegraphics[width=0.48\textwidth]{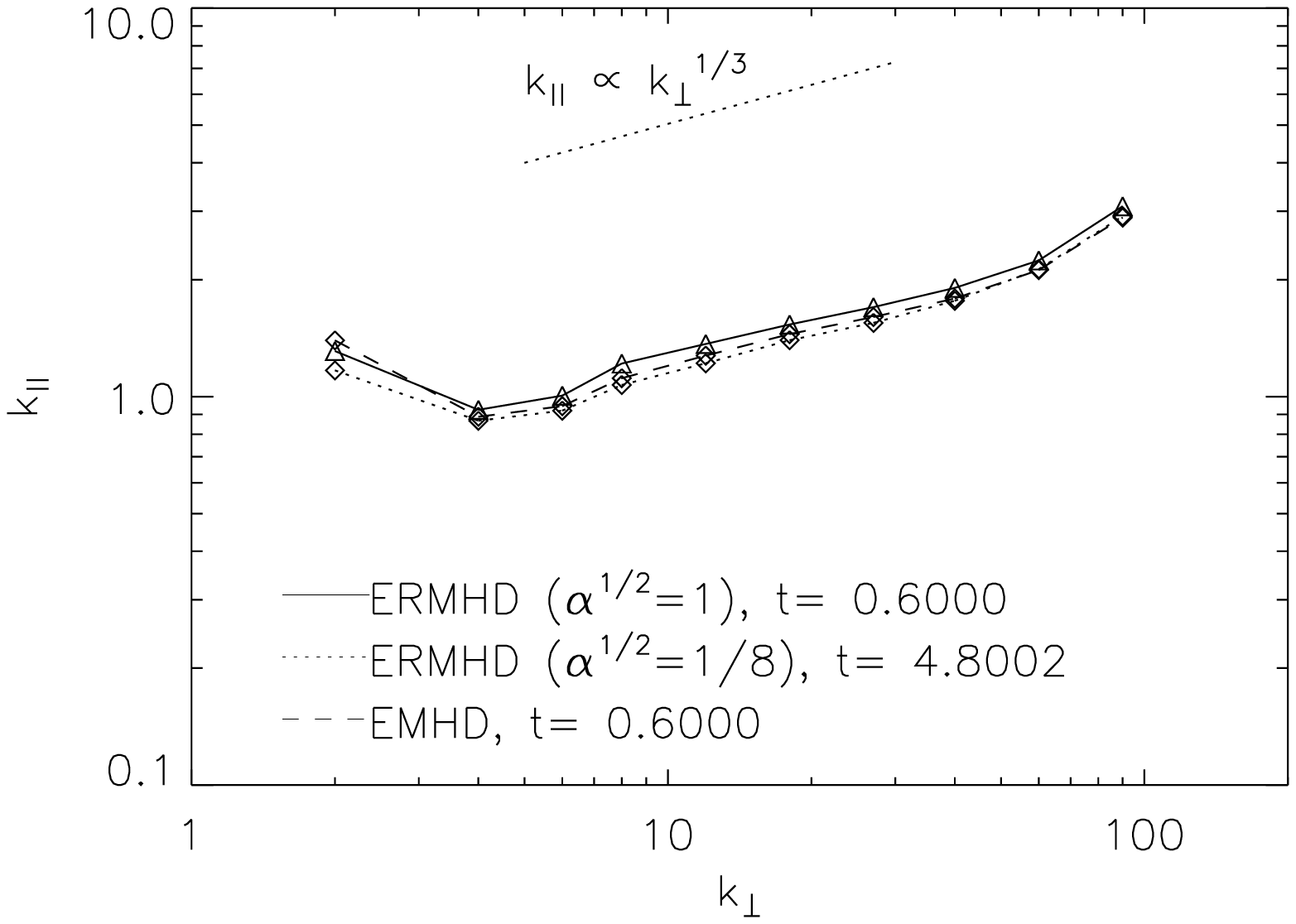}  
\caption{ Comparison of EMHD (Run E256D-EL) and ERMHD (Runs ER256D1-EL and ER256D8-EL).
   {\it Left}: energy spectra and time evolution of magnetic energy density (inset). 
   The energy spectra are taken
   at the time marked by the arrow in the inset. Note that the horizontal axis
   of the inset is $\sqrt{\alpha}t$, where $\alpha$ is a parameter that appears
   in the ERMHD formalism (Eq.~[\ref{eq_alpha}]). 
   The vertical axis of the inset is $b^2$ (or, in case of ER256D8-EL, $\tilde{b}^2$).
   {\it Right}: comparison of anisotropy. The calculations are done 
   at the time marked by the arrow in the inset of left panel.
\label{fig:e_n_ermhd}
}
\end{figure*}

\section{Numerical Methods}
As we described earlier, EMHD and ERMHD equations involve with
time evolution of magnetic field only.
Since magnetic field is divergence-free,
we can use incompressible numerical schemes to solve the equations.

\subsection{The EMHD code}
We adopt a pseudospectral code to solve the normalized
EMHD equation in a periodic box of size $2\pi$:
\begin{equation}
\frac{\partial {\bf B}}{\partial t}=-
     \nabla \times \left[
    (\nabla \times {\bf B}) \times{\bf B} \right] 
    + \eta^{\prime} \nabla^{2} {\bf B},
     \label{beq}
\end{equation}
where magnetic field, time, and length are normalized by
a mean field $B_0$, the whistler time $t_w=L^2(\omega_{pe}/c)^2/\Omega_e$
($\Omega_e$= electron gyro frequency), and
a characteristic length scale $L$ 
(see, for example, Galtier \& Bhattacharjee 2003).
The resistivity $\eta^{\prime}$ in equation (\ref{beq}) is dimensionless.
The dispersion relation of a whistler waves in this normalized units
is $\omega=kk_{\|}B_0$.  
The magnetic field consists of the uniform background field and a
fluctuating field: ${\bf B}= {\bf B}_0 + {\bf b}$.
The strength of
the uniform background field, $B_0$, is set to 1.
We use up to $512^3$ collocation points.
At $t=0$, the random magnetic field is restricted to the range
$2\leq k < 5$ in wavevector space.
The amplitudes of the random magnetic field at $t=0$ is $\sim 1.2$.
In order to have a more extended inertial range, we
use hyperdiffusivity for the diffusion terms\footnote{
   We do not observe a strong bottleneck effect, which is a common
   feature in numerical
   hydrodynamic simulations with hyperviscosity/hyperdiffusivity. 
   Therefore expect that hyperdiffusivity does not substantially alter physics 
   in the inertial range.
   However, this should be clarified in the future.
   Discussions on the bottleneck effect for 2-dimensional EMHD can be found in
   Biskamp, Schwarz, \& Celani (1998).
}.
The power of hyperdiffusivity
is set to 3 for all simulations, so that the dissipation term in the above equation
is replaced with
  $\eta_3 (\nabla^2)^3 {\bf B}$.
We perform both driven and decaying turbulence simulations.

\subsection{The ERMHD Code}
The ERMHD equation is only slightly different from the EMHD equation.
Therefore, we also adopt a pseudospectral code to solve the normalized
ERMHD equation in a rectangular periodic box of size $2\pi \times 2\pi \times 6\pi$:
\be
\frac{\partial \tilde{\bf B}}{\partial \tilde{t}}=-
   \nabla_{\perp} \times ({\bf B} \cdot \nabla \tilde{\bf B})
    + \eta^{\prime} \nabla^{2} \tilde{\bf B},
     \label{beq_er}
\ee
where  
\bea
     \tilde{\bf B} \equiv {\bf B}_0 + \tilde{\bf b}, \\
     \tilde{\bf b}= {\bf b}_{\perp}+\sqrt{\frac{1}{\alpha}} b_{\|} \hat{\bf z}, \\
     \tilde{t} \equiv \sqrt{\alpha} t,
\eea
where $\eta^\prime$ is dimensionless resistivity, 
${\bf b}_{\perp}$ and $b_{\|}\hat{\bf z}$ are
perpendicular and parallel components of magnetic field, respectively,
and
definition of $\alpha$ is given in Eq.~(\ref{eq_alpha}).
Note that the ERMHD equation is very similar to the EMHD equation.
Numerical setup is therefore similar to that of the EMHD case, 
except the shape of the simulation box.

We use an elongated numerical box for ERMHD to satisfy
the condition $k_{\perp} \gg k_{\|}$. 
The numerical box is  3 times longer in the direction of the mean magnetic field.
The wavenumbers along the mean field direction have fractional values 
\be
  k_{\|} = 1/3, 2/3, 1, 4/3, ...,
\ee
while those of perpendicular direction have integer values\footnote{
 Or, equivalently, we may use $k_{\|}=1,2,3, ...$ and 
     $k_{\perp}=3,6,9, ...$. In this case, the size of the computational box is
     $(2\pi/3) \times (2\pi/3) \times (2\pi)$, where $2\pi$ is the side
     along the mean magnetic field.
}.
We perform only decaying turbulence simulations.
At t=0, Fourier modes with $1/3\le k_{\|} \le 1$ and 
$4.5/\sqrt{2} \le k_{\perp} \le 4.5\sqrt{2}$
are excited.
The strength of the mean magnetic field ($B_0$) is $1$
and the rms value of the random magnetic field (actually $\tilde{b}$) at t=0 is $\sim 0.071$.
Therefore, the critical balance, $bk_{\perp}/B_0 k_{\|} =1$, is roughly satisfied at t=0.
For comparison, we perform a similar EMHD simulation (Run E256D-EL).

\section{Results: EMHD and ERMHD}
We compare EMHD and ERMHD turbulence in Fig.~\ref{fig:e_n_ermhd}.
Left panel of Fig.~\ref{fig:e_n_ermhd} shows time evolution and energy spectra of 
decaying EMHD and ERMHD runs (see Runs E256D-EL, ER256D1-EL, and ER256D8-EL in 
Table 1).
The inset shows time evolution of magnetic energy density.
The solid curve is for ERMHD with $\sqrt{\alpha}=1$, dotted curve for ERMHD with
$\sqrt{\alpha}=1/8$, and the dashed curve for EMHD.
The horizontal axis represents time multiplied by $\sqrt{\alpha}$ and
the vertical axis twice the magnetic energy density.
In case of ERMHD with $\sqrt{\alpha}=1/8$ (i.e.~Run ER256D8-EL), the vertical axis
is not $b^2$, but $\tilde{b}^2 = b_\perp^2 + b_{\|}^2/\alpha$.
All three curves follow a similar decay law.
The main plot of Fig.~\ref{fig:e_n_ermhd} shows energy spectra at $\sqrt{\alpha}t=0.6$
(see the arrow in the inset).
All three spectra are consistent with the expected $k^{-7/3}$ spectrum.

We compare anisotropy of EMHD and ERMHD in right panel of Fig.~\ref{fig:e_n_ermhd}. 
We obtain the plot at $\sqrt{\alpha}t=0.6$.
All three curves show similar scaling relations, $k_{\|} \propto k_\perp^{1/3}$.
Note however that, although we do not show it in this paper, anisotropy tends to get
stronger as turbulence decays further. 
We will describe the method of obtaining anisotropy and its limitations in Section 
\S\ref{sect_ani1}.

To summarize, we observe that both EMHD and ERMHD turbulence exhibit similar spectra
and anisotropy.
Spectra and time evolution of magnetic energy density of
ERMHD turbulence seem to be invariant under the following transformation,
\bea
   {\bf B} \rightarrow \tilde{\bf B}, \mbox{~~~~~} t \rightarrow \sqrt{\alpha} t.
\eea

\begin{figure*}[h!t]
\includegraphics[width=0.48\textwidth]{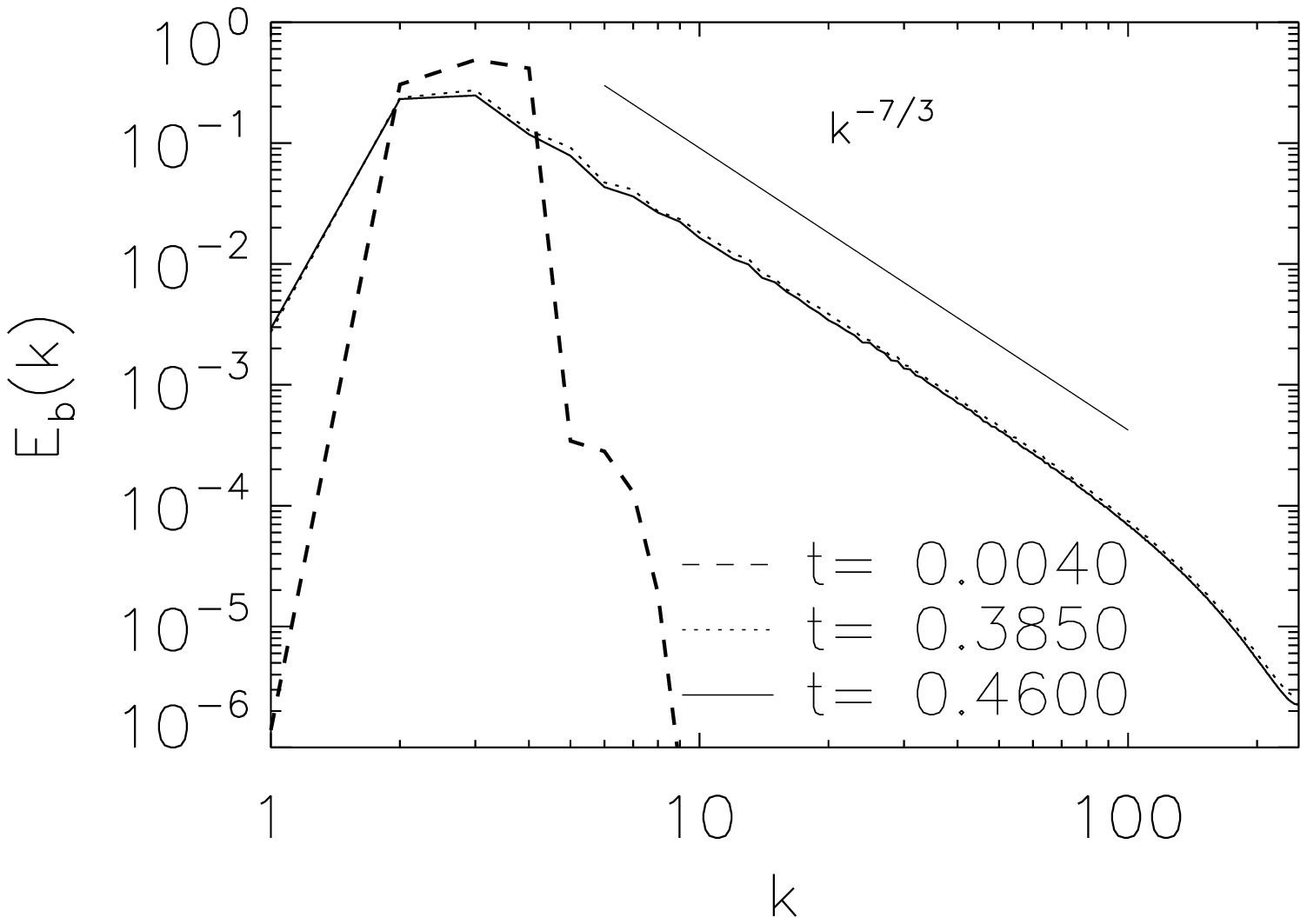}  
\includegraphics[width=0.48\textwidth]{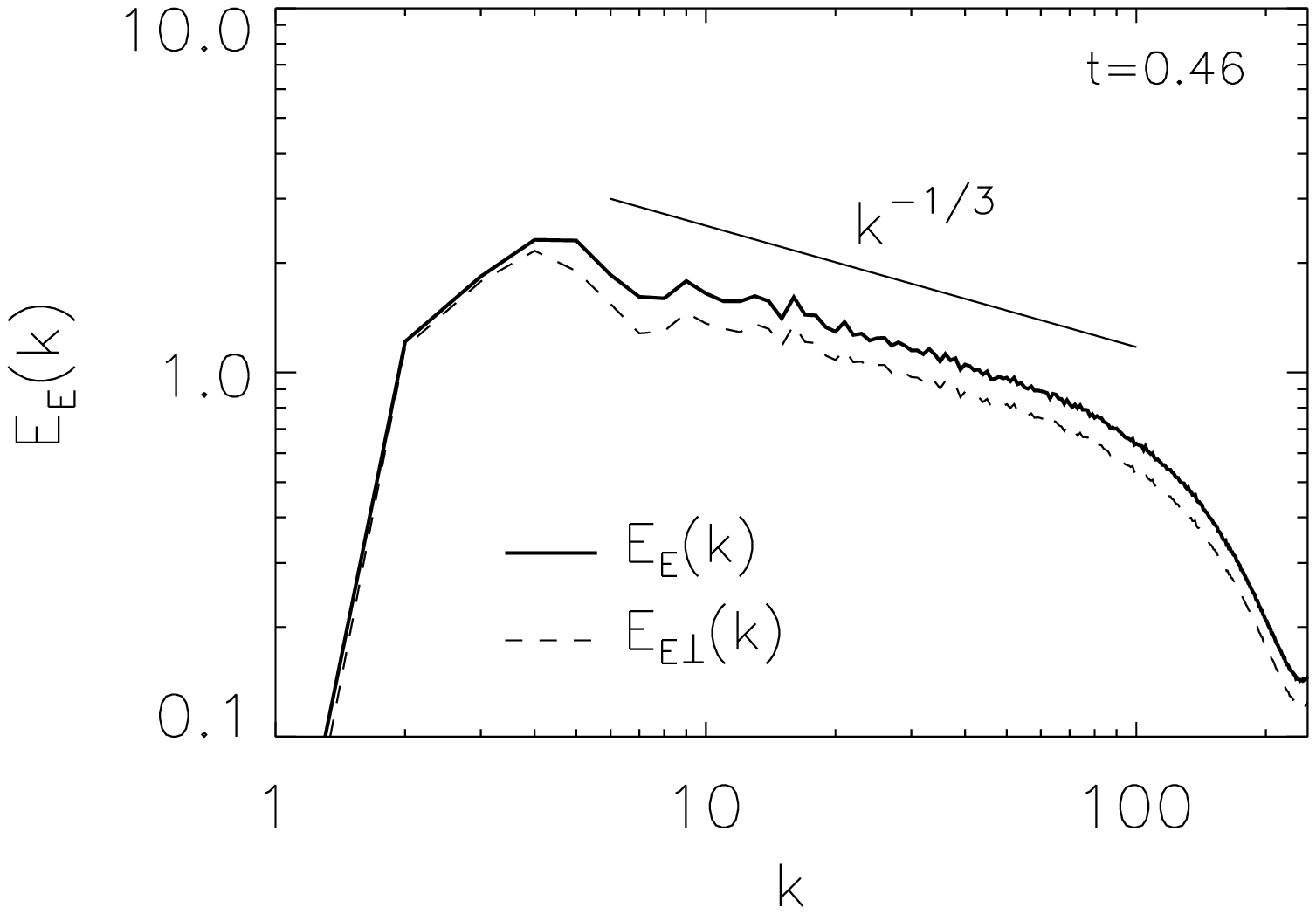}  
\caption{ Spectra of ${\bf B}$ (left) and ${\bf E}$ (right) of EMHD turbulence (Run E512D).
   {\it Left}: spectrum at different time points.
   At $t\approx 0$, only small-wavenumber Fourier modes are excited.
   At later times, energy gradually cascades down to small scales (i.e.~ large $k$ regions)
   and a $k^{-7/3}$ inertial range develops.
   {\it Right}: spectrum of electric field ${\bf E}$ at t=0.46.
   The solid curve is the spectrum of total electric field and the dashed curve is that of
   electric field perpendicular to the mean field ${\bf B}_0$.
\label{fig:spect}
}
\end{figure*}

\section{Results: Scaling of EMHD} \label{sect_512}
In the previous section, we have studied turbulence in a rectangular box elongated 
along the mean field direction.
Note that the numerical resolution in the perpendicular direction is $256\times 256$,
which limits the inertial range of turbulence. 
To increase the inertial range we need to increase numerical resolution 
in the perpendicular direction.
Studying ERMHD is more difficult than studying EMHD because 
we need to use an elongated numerical box for ERMHD to satisfy the condition
$k_{\perp} \gg k_{\|}$.
For EMHD, there is no necessity of using such an elongated numerical box.
Therefore, EMHD is more suitable for high resolution simulations.
Since EMHD turbulence and ERMHD turbulence
exhibit similar scaling relations, 
we only consider a high resolution EMHD simulation in this section. 

We perform a decaying EMHD simulation with $512^3$ grid points (Run E512D in Table 1).
At t=0, Fourier modes with $2\leq k <5$ are excited. 
Here $k =\sqrt{k_\perp^2 +k_{\|}^2}$ and, thus,
the initial perturbation is isotropic (i.e.~$k_\| \sim k_{\perp}$). 
The strength of the mean magnetic field ($B_0$) is $1$
and the rms value of the random magnetic field ($b$) is $\sim 1.21$.
Therefore, the critical balance, $bk_{\perp}/B_0 k_{\|} =1$, is roughly satisfied at t=0.

In this section, we focus on anisotropy of EMHD turbulence.
CL04 showed that anisotropy of EMHD turbulence is consistent
with $k_{\|}\propto k_{\perp}^{1/3}$.
However, since it is difficult to measure anisotropy of EMHD turbulence,
more careful assessments
of anisotropy are necessary.
In this subsection, we develop and test new techniques for anisotropy.
We analyze anisotropy of turbulence after turbulence has developed a full inertial range.

The energy spectra in left panel of Figure~\ref{fig:spect} show how the inertial range develops.
At $t=0$ only large-scale 
(i.e. small $k$)
Fourier modes are excited. The dashed curve in Figure~\ref{fig:spect} shows the
initial spectrum.
As the turbulence decays, the initial energy cascades down to small scales 
and, as a result, small scale (i.e. large $k$) modes are excited.
When the energy reaches the dissipation scale at $k\gtrsim100$,
the energy spectrum goes down without changing its slope (the dotted and the 
solid curves).
The slope at this stage is
very close to that of the predicted spectrum:
\begin{equation}
  E(k) \propto k^{-7/3}.
\end{equation}
For the study of anisotropy, we use the data cube at $t\approx 0.46$.
The solid curve in left panel of Figure~\ref{fig:spect} represents the spectrum at that time.

The right panel of Figure~\ref{fig:spect} shows spectra of electric field.
Both the total electric field (solid curve) and its xy-component (dashed curve)
show spectra compatible with $k^{-1/3}$. (Note that we assume ${\bf B}_0=B_0 \hat{\bf z}$.)
This result is consistent with the ones obtained by gyrokinetic simulations (Howes et al. 2008a)
and Hall MHD simulations (see Dmitruk \& Matthaeus 2006; Matthaeus et al. 2008).
Bale et al. (2005) reported a similar electric fluctuation spectrum in the solar wind.

\begin{figure*}[h!t]
\includegraphics[width=0.48\textwidth]{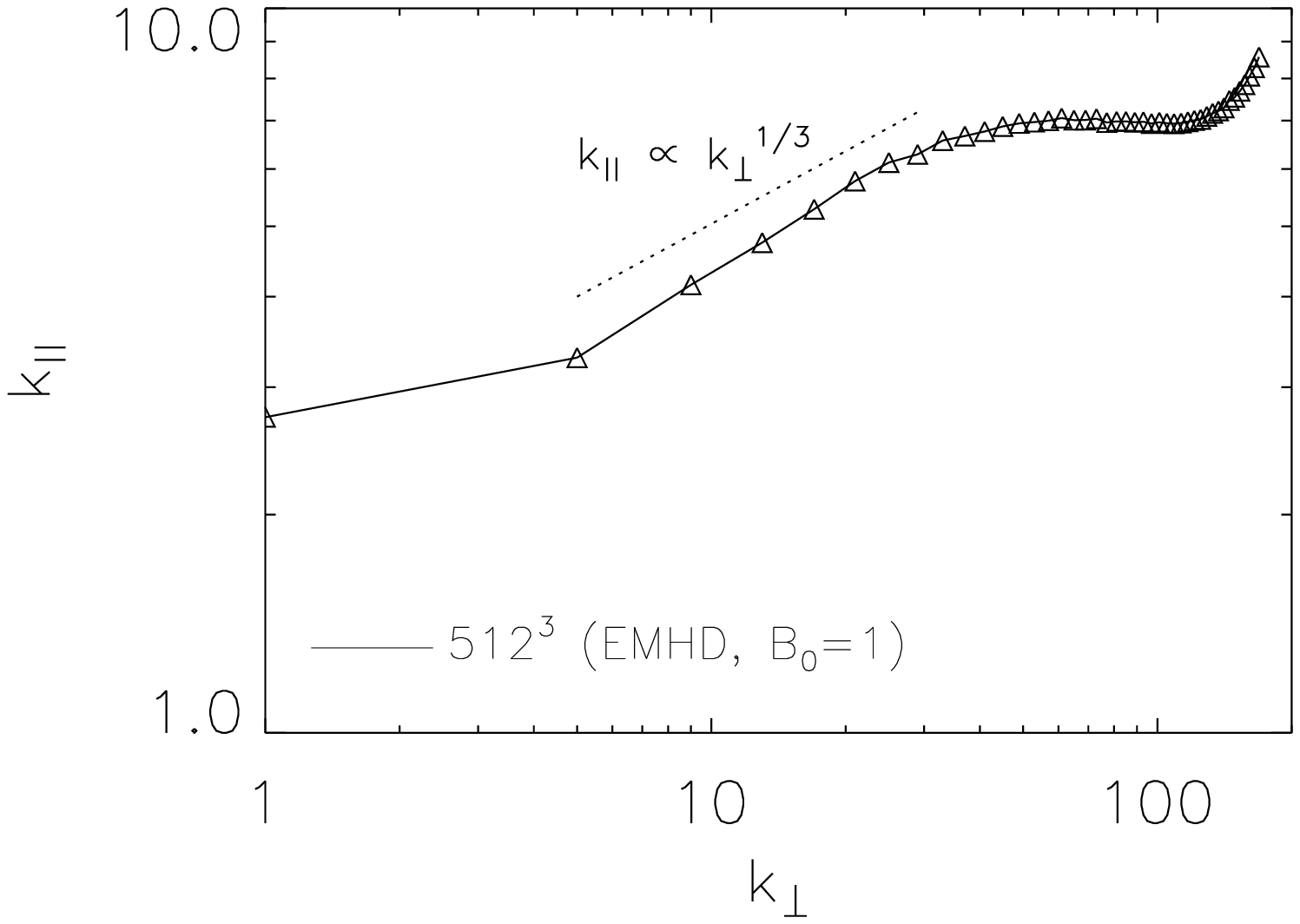}  
\includegraphics[width=0.48\textwidth]{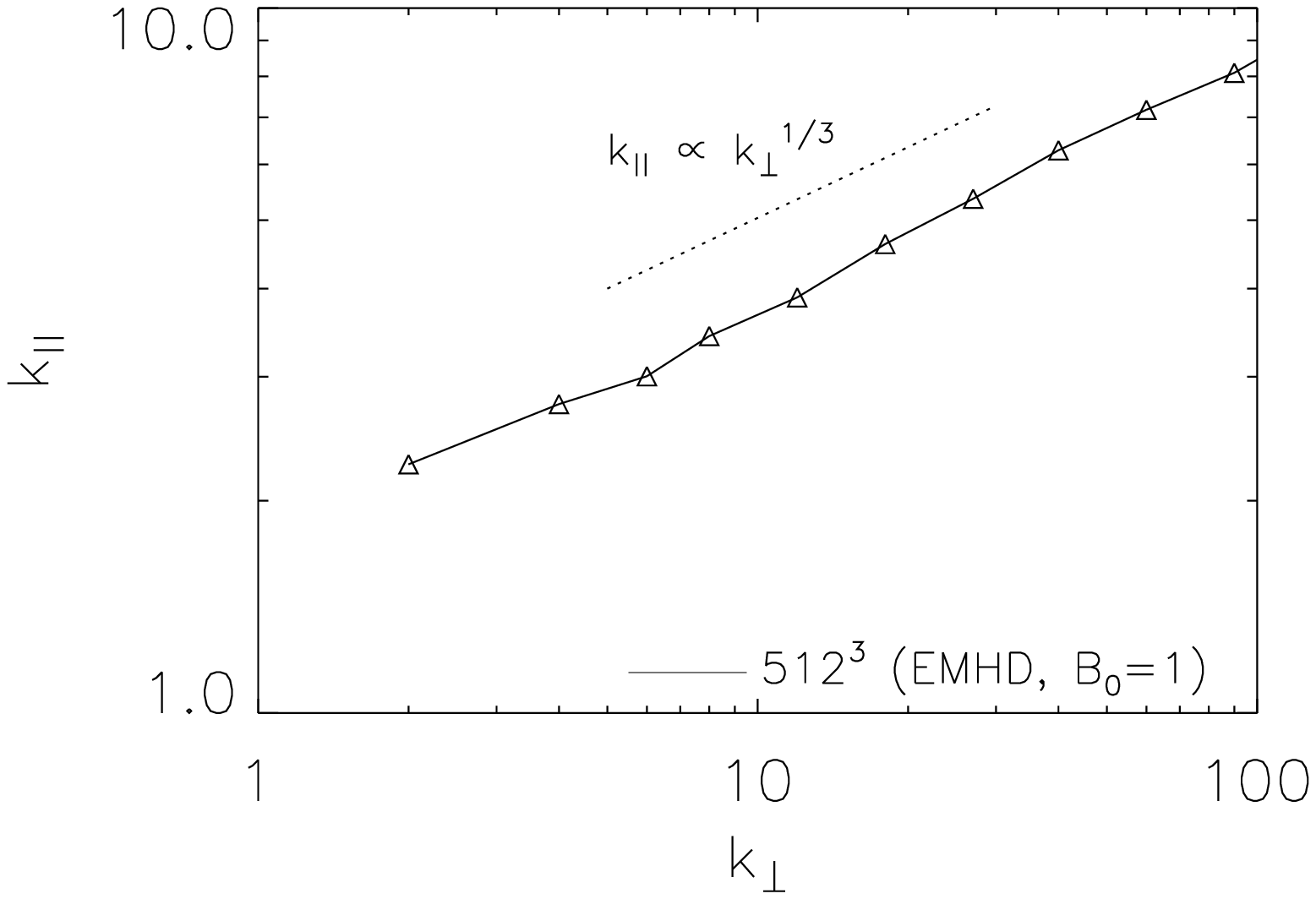}  
\caption{ Anisotropy of EMHD turbulence (Run E512D).
 {\it Left}: the old method (Eq.~[\ref{kparLl}]) is used.
 {\it Right}: a new method (Eq.~[\ref{eq:newmethod}]) is used.
\label{fig:kpar_oldnew}
}
\end{figure*}

\subsection{Anisotropy: method 1} \label{sect_ani1}
In CL04, we noted that
the quantity 
${\bf B}_L\cdot \nabla {\bf b}_l$
is proportional to $B_L k_{\|} b_l$, where ${\bf B}_L$ is the local mean field,
${\bf b}_l$ the fluctuating field at scale $l$, and
$k_{\|}$ the wave number 
parallel to the local mean magnetic field.
We obtained ${\bf B}_L$ by
eliminating
Fourier modes whose perpendicular wavenumber is greater than $k/2$.
We obtained the fluctuating field ${\bf b}_l$ by eliminating
Fourier modes  whose perpendicular wavenumber is less than $k/2$.
In CL04, we calculated anisotropy in Fourier space:
\begin{equation}
k_{\|}(k_{\perp}) \approx \left(
     \frac{\sum_{k\leq |{\bf k}^{\prime}| <k+1}
   |\widehat{ {\bf B}_L\cdot \nabla {\bf b}_l   }|_{{\bf k}^{\prime}}^2  }
  { B_L^2 \sum_{k\leq |{\bf k}^{\prime}| <k+1}
                     |\hat{\bf b}|^2_{{\bf k}^{\prime}} }
               \right)^{1/2}.    \label{kparLl}
\end{equation}
The reason why we did the calculation in Fourier space is that
there might be unknown contamination from scales 
other than $l$.  
We plot anisotropy obtained this way in left panel of
Fig.~\ref{fig:kpar_oldnew}.

We note that, when we obtain 
the fluctuating field ${\bf b}_l$ by filtering out
Fourier modes  whose perpendicular wavenumber is less than $k/2$ {\it or greater than}
$2k$, we may calculate anisotropy directly in real space:
\be
    {\bf B}_L \cdot \nabla {\bf b}_l \approx B_L k_{\|} b_l 
    \rightarrow k_{\|} \approx 
    \left(
           \frac{  <|{\bf B}_L \cdot \nabla {\bf b}_l|^2 > }
                { B_L^2 b_l^2 } 
     \right)^{1/2} .
   \label{eq:newmethod}
\ee
We plot anisotropy obtained this way in right panel of
Fig.~\ref{fig:kpar_oldnew}.
The result is compatible with the 
expected anisotropy, $k_{\|}\propto k_{\perp}^{1/3}$.
The limitation of this method is that
the choice of the filtering wavenumbers, $k/2$ and $2k$, 
is an arbitrary one.
When we use filtering wavenumbers of $k/\sqrt{2}$ and $\sqrt{2}k$,
we get a slightly weaker anisotropy.

\begin{figure*}[h!t]
\includegraphics[width=0.48\textwidth]{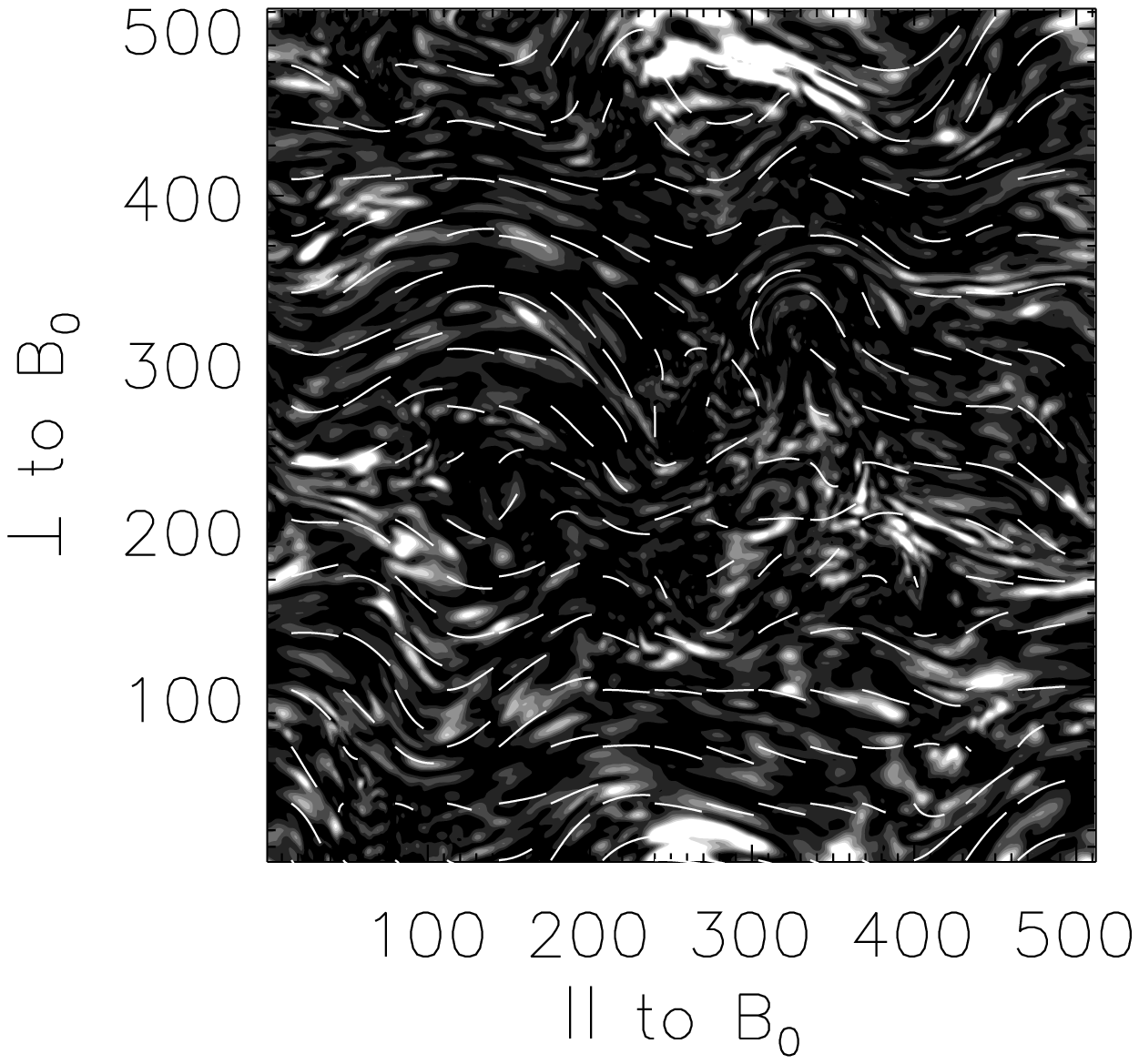}  
\includegraphics[width=0.48\textwidth]{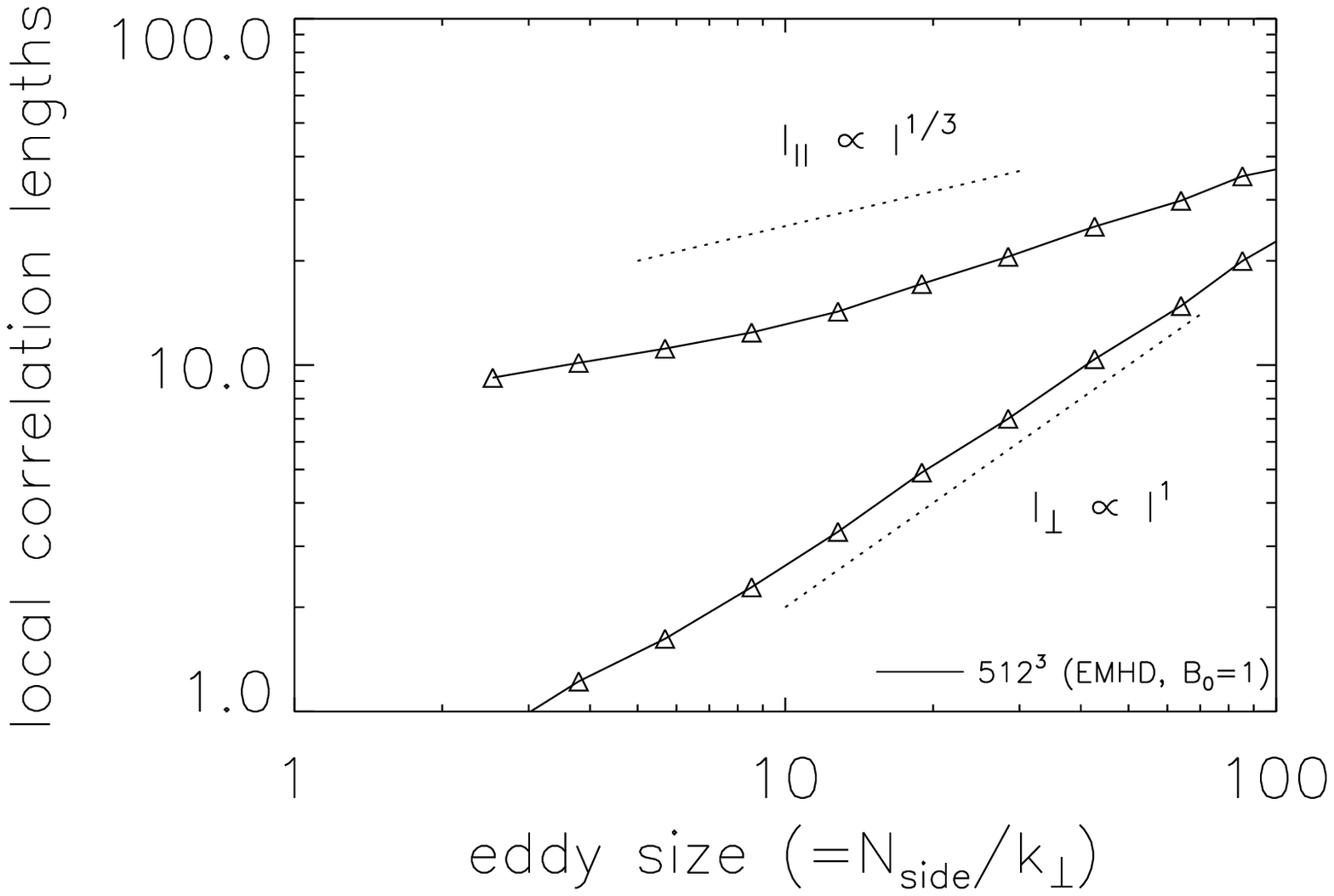}  
\caption{ Anisotropy based on local correlation lengths (Run E512D).
 {\it Left}: visualization. White lines denote 
 {\it local} mean magnetic field ${\bf B}_L$ obtained by filtering out large-$k$ Fourier modes.
 Contours show structure of small-scale eddies obtained by
 retaining Fourier modes near the scale of interest and filtering out all other
 scale Fourier modes. The global mean field ${\bf B}_0$ is parallel to the horizontal axis.
 {\it Right}: a new method based on correlation lengths in local frame.
 The parallel correlation length $l_{\|}$, for example, is the average of the 
  correlation length
 measured along the local mean magnetic field direction at each point. 
 (Note that the local mean magnetic field directions are marked
  by white lines in the left panel).
 In Run E512D, $N_{side}=512$.
\label{fig:kpar_corr}
}
\end{figure*}

\subsection{Anisotropy: method 2}
In the previous subsection, we assumed that
${\bf B}_L\cdot \nabla {\bf b}_l \propto B_L k_{\|} b_l$.
Is this really correct?
If small-scale eddies are aligned with local
mean field, the above assumption
should be true.
However, if the major axes of small-scale eddies are substantially misaligned with
local large-scale mean field, the quantity ${\bf B}_L\cdot \nabla {\bf b}_l$
may sample perpendicular wavenumber of the eddies.
In this subsection, we investigate alignment of small-scale eddies with respect to
local mean magnetic field.

We first visualize the alignment effect.
Left panel of Fig.~\ref{fig:kpar_corr} shows a snapshot of 
magnetic field structure in a plane parallel to the {\it global} mean magnetic field.
The global mean magnetic field is parallel to the horizontal axis in the Figure. 
The white lines in Figure represent local mean magnetic field.
We obtain the local mean field
by filtering out Fourier modes with $k>15$.
The contours represent magnetic energy density of small-scale field.  
We obtain the small-scale field
by filtering out Fourier modes with $k\leq 15$ or $k\geq 60$.
Since the alignment effect, if any, should be 3-dimensional effect in nature,
we may not be able to visualize the effect correctly in a 2-dimensional
plane.
Nevertheless, we can observe that small-scale eddies are well aligned with
the local mean field directions.

If small-scale eddies are well aligned with the local mean field,
the correlation length along the local mean field
should be larger than that of perpendicular directions.
Furthermore, when the alignment effect is present, the parallel correlation length
should follow a $l_{\perp}^{1/3}$ (or, $k_{\perp}^{-1/3}$) power law.
We investigate the behavior of parallel and perpendicular correlation lengths
by changing the scale $l$.\footnote{
  We use a similar filtering method as in the previous subsection.
  We obtain the local mean field, ${\bf B}_L$, by
  eliminating
  Fourier modes whose perpendicular wavenumber is greater than $k/2$.
  We obtain the fluctuating field, ${\bf b}_l$, by eliminating
  Fourier modes  whose perpendicular wavenumber is less than $k/2$ or greater than $2k$.
  Note that $l= 2\pi/k$, or in terms of grid units $l= N_{side}/k$, where $N_{side}=512$
  in Run E512D.
}
We define the correlation lengths as the distances at which the 2-point
correlation function $\langle {\bf b}_l({\bf r}_1)\cdot{\bf b}_l({\bf r}_2) \rangle$ drops by 50\%.
The parallel correlation length is measured along the directions of
local mean field, ${\bf B}_L$, and the perpendicular one for perpendicular directions.
In right panel of Fig.~\ref{fig:kpar_corr}, we plot the parallel  and
perpendicular
correlation lengths as functions of small-scale eddy size $l\propto 1/k$.
The parallel correlation lengths (upper curve) is of course longer than
the perpendicular ones (lower curve).
On average, the parallel correlation lengths seem to follow the expected $l^{1/3}$ scaling. 
However, the slope is shallower than $1/3$ for small-size eddies and 
steeper than $1/3$ for large-size eddies.
Roughly speaking, the perpendicular correlation lengths follow $l^1$ scaling, which
is reasonable.

All in all, although it may not be not a perfect method, 
the filtering method reasonably reveals anisotropy of EMHD turbulence.
The anisotropy of EMHD turbulence revealed by the filtering process is
consistent with the $k_{\|} \propto k_{\perp}^{1/3}$ scaling.

\begin{figure*}[h!t]
\includegraphics[width=0.32\textwidth]{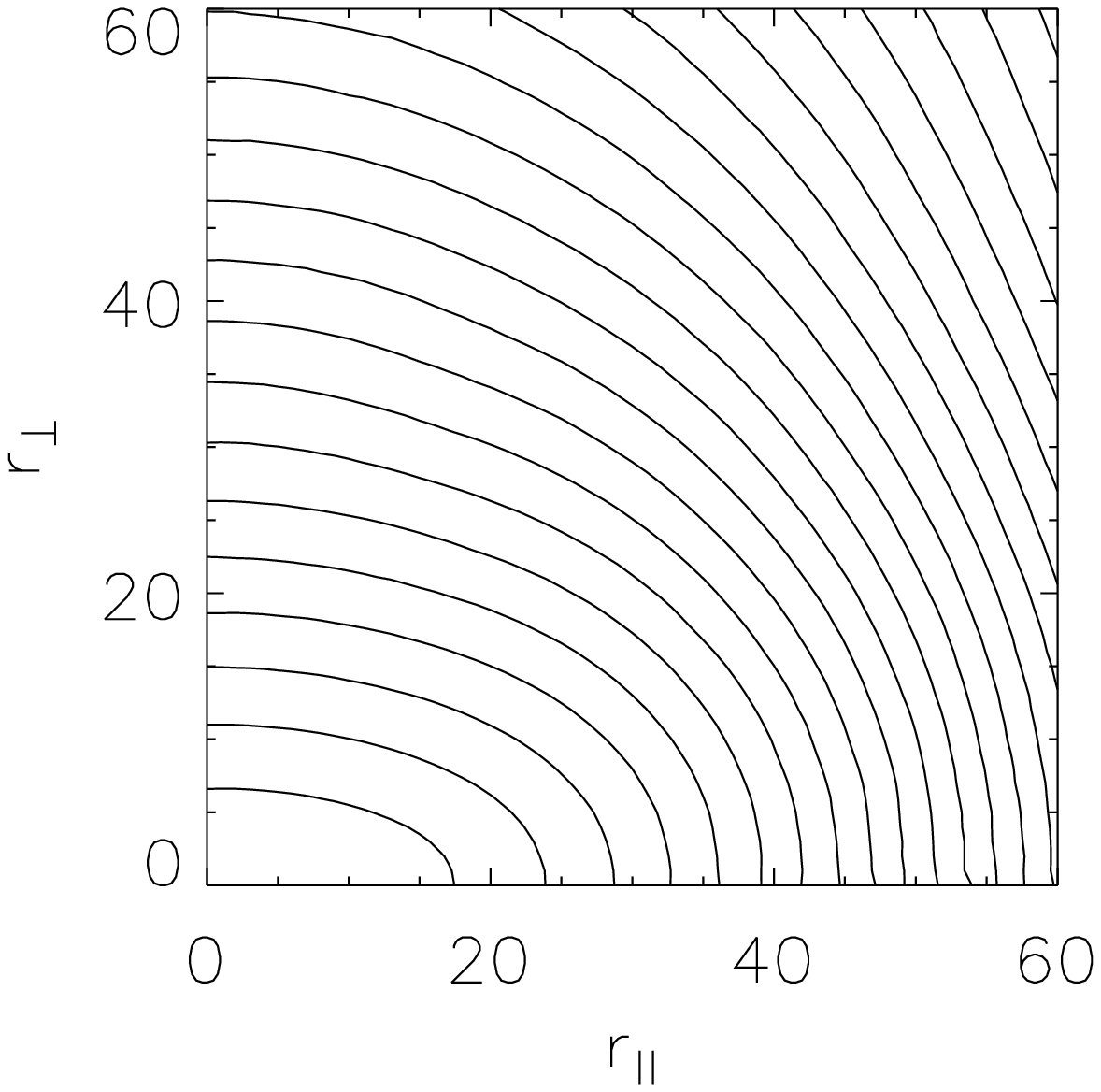}  
\includegraphics[width=0.32\textwidth]{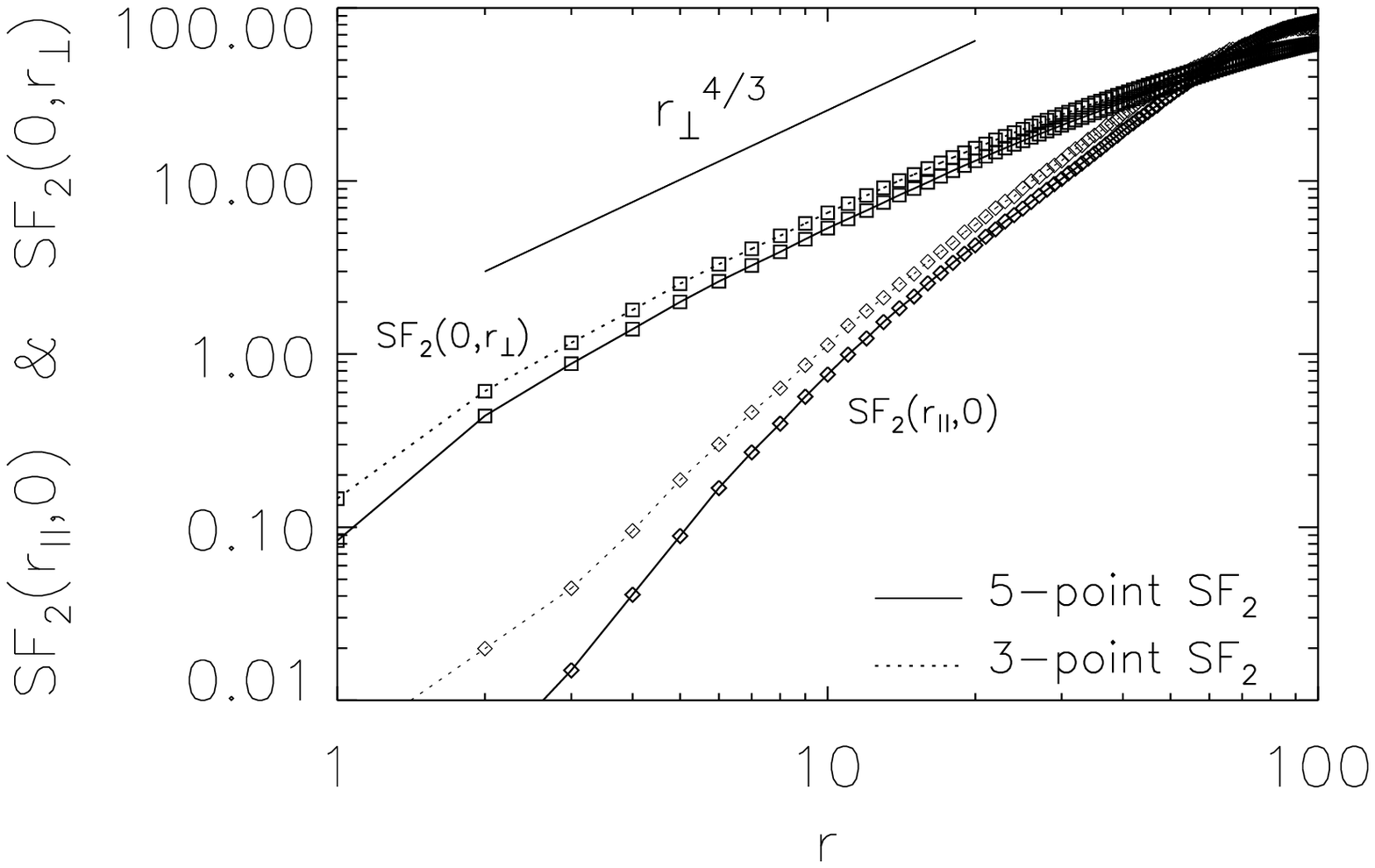}  
\includegraphics[width=0.32\textwidth]{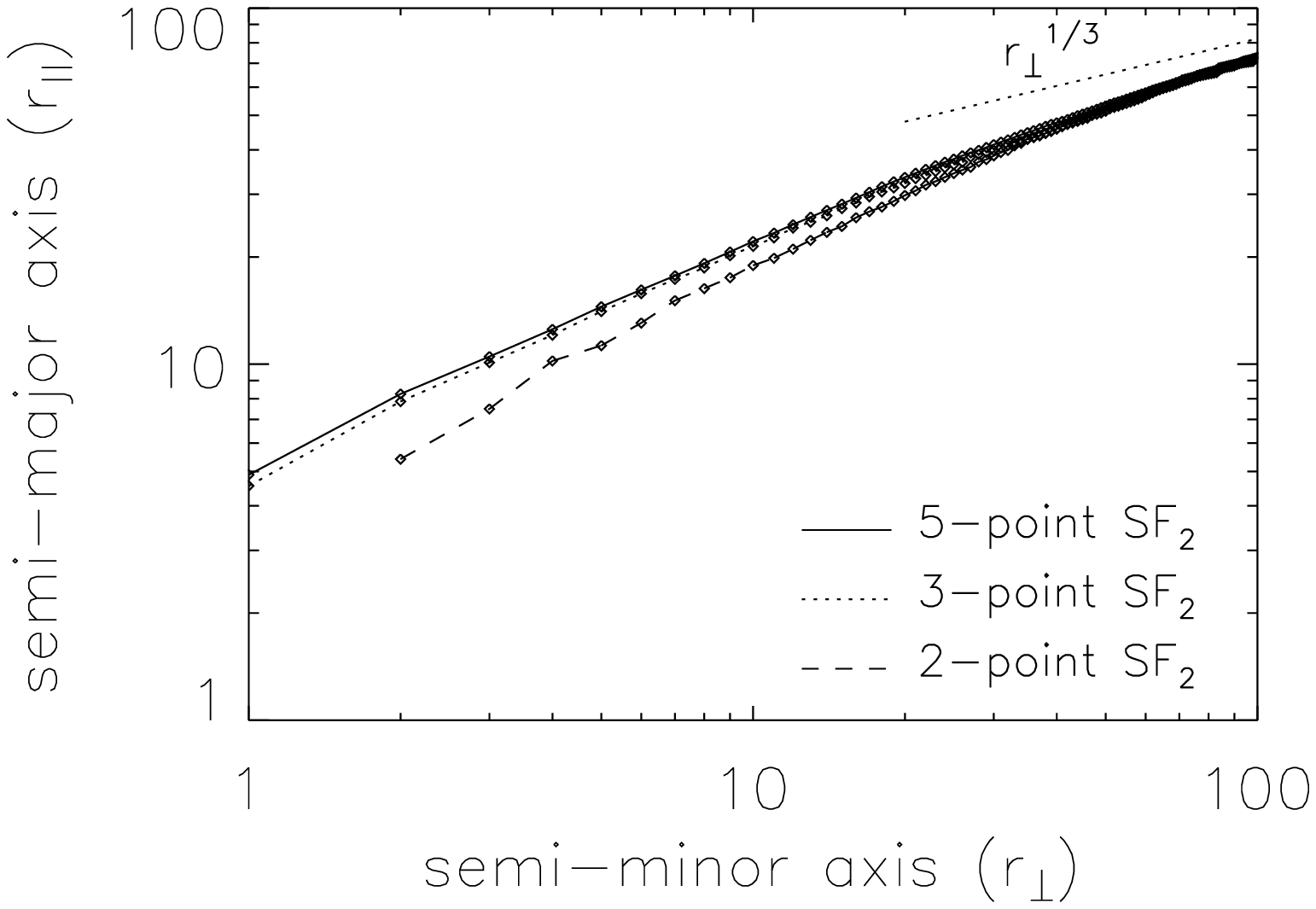}  
\caption{Anisotropy from the second-order structure function (Run E512D).
   {\it Left}: contour diagram obtained from the 5-point second-order structure function.
   {\it Middle}: comparison of the 3-point SF$_2$ and the 5-point  SF$_2$.
     The parallel structure function $SF_2(r_{\|},0)$, for example,
     can be obtained from values of $SF_2$ on the horizontal axis of the contour diagram.
   {\it Right}: the relation between the semi-minor axis ($\propto 1/k_{\perp}$)
     and the semi-major axis ($\propto 1/k_{\|}$) of the contours.
\label{fig:conto_kpar}
}
\end{figure*}

\subsection{Anisotropy: Multi-point second-order structure functions} \label{sect:sf2}
We may visualize scale-dependent anisotropy using the second-order structure function
calculated in the local frame, 
which is aligned with local mean magnetic field
${\bf B_L}$:
\begin{equation}
    \mbox{SF}_2(r_{\|},r_{\perp})=<|{\bf B}({\bf x}+{\bf r}) -
                 {\bf B}({\bf x})|^2>_{avg.~over~{\bf x}},
\end{equation}
where ${\bf r}=r_{\|} {\hat {\bf r}}_{\|} +r_{\perp} {\hat {\bf r}}_{\perp}$.
The vectors
${\hat {\bf r}}_{\|}$ and ${\hat {\bf r}}_{\perp}$ are unit vectors
parallel and perpendicular to the local mean field ${\bf B_L}$, respectively.
See Cho \& Vishniac (2000) for the 
detailed discussion of the local frame.
Left panel of Fig.~\ref{fig:conto_kpar} is an example of such visualization\footnote{
     Note however that we do not use the usual 2-point second-order structure 
     function for the plot. Instead, we use the 5-point second-order structure function
     that will be discussed later in this subsection.
}.
The horizontal axis corresponds to the local mean field directions.
The shapes of contours illustrate scale-dependent anisotropy:
the large eddies are roughly isotropic and smaller eddies are more elongated.
However, we should be careful when we derive a quantitative scaling relation for anisotropy
from the contour diagram.

The 2-point second-order structure function of a variable $A$
\be
   SF_2(r) = (\delta A_r)^2=<|  A({\bf x}+{\bf r})-A({\bf x})|^2 >
\ee
is in general related to the energy spectrum of the variable, $E_A(k)$,
through 
\be
   SF_2(r)=(\delta A_r)^2 \sim kE_A(k),  \label{eq:sf2kek}
\ee
where $k\propto 1/r$ is the wavenumber.
Therefore, when $E_A(k)\propto k^{-m}$, we have 
\be
   SF_2(r) \propto r^{m-1}.
\ee
For example,
in fully developed Kolmogorov turbulence ($E(k)\propto k^{-5/3}$), 
the scaling relation of the second-order longitudinal structure function is given by
\be
 SF_2(r)=(\delta v_r)^2 \sim kk^{-5/3} \propto k^{-2/3} \propto r^{2/3}.
\ee

However, when the slope of the turbulence spectrum is 
    steeper than $k^{-3}$, the relation in Eq.~(\ref{eq:sf2kek})
    becomes invalid and 
    $SF_2(r) \propto r^{2}$ regardless
    of the slope of the energy spectrum (see Appendix B).
In EMHD, the energy spectrum expressed in terms of $k_{\perp}$ scales as
$E(k_\perp)\propto k_{\perp}^{-7/3}$ 
and we expect that the second-order structure function
scales as 
\be
   SF_2(0,r_\perp)\propto r_{\perp}^{4/3}.
\ee
On the other hand, the energy spectrum expressed in terms of $k_{\|}$ is expected to be
$E(k_\|)\propto k_{\|}^{-5}$ when anisotropy scales as $k_{\|}\propto k_{\perp}^{1/3}$
(see CL04).
Therefore, the one-to-one correspondence between the second-order structure function 
and the energy spectrum is not valid for parallel direction.
This means we cannot use the second-order structure function to reveal
true anisotropy.

We have shown that the 2-point second-order structure function is not suitable
for quantitative study of anisotropy in EMHD turbulence. However, it is possible
to construct multi-point second-order structure functions that
can be used for variables with steep energy spectra 
(see Falcon, Fauve, \& Laroche 2007; Lazarian \& Pogosyan 2008;
see also Appendix C).
The 3-point $SF_2$ was used by Falcon et al. (2007) and Lazarian \& Pogosyan (2008).
The 3-point $SF_2$ will work for energy spectrum as steep as $\sim k^{-5}$.
In Appendix C, we discuss how to construct 4-point and 5-point structure functions.
The 5-point $SF_2$ works for energy spectrum as steep as $\sim k^{-9}$.

Left panel of Fig.~\ref{fig:conto_kpar} is obtained with this 5-point structure function.
Middle panel of the Figure shows the parallel structure functions, $SF_2(r_{\|},0)$, and 
the perpendicular structure functions, $SF_2(0,r_{\perp})$.
We plot the results of 3-point (dotted lines) and 5-point (solid lines) structure functions.
In the perpendicular direction, both structure functions reasonably follow the 
expected scaling of $r_{\perp}^{4/3}$.
In the parallel directions, however, 
we observe that, when $r_{\|}\gtrsim 10$, the structure functions 
are much shallower than the expected scaling of $r_{\|}^{4}$, which is from
$SF_2 \propto k_{\|} E(k_{\|}) \propto k_{\|}^{-4} \propto r_{\|}^4$.
In the parallel directions, the 5-point structure function is steeper than 
the 3-point one.

One should note that it is very difficult to obtain a well-defined power-law scaling
for the parallel direction.
This is because the inertial range in the parallel direction, if any, is extremely short.
In the perpendicular direction, the inertial range spans from the outer scale ($k_{\perp}\sim 3$)
to the dissipation scale ($k_\perp \sim 100$).
Suppose that $k_{\|}\propto k_{\perp}^{1/3}$ is the true anisotropy.
Then, in the parallel direction, the inertial range spans
{}from $k_{\|}\sim 3$ to $k_{\|}\sim 3\times (100/3)^{1/3}\sim 10$.
Therefore, it is virtually hopeless to reveal true anisotropy from contour diagram.
Indeed, right panel of Fig.~\ref{fig:conto_kpar} shows that the anisotropy
derived from the relation between semi-major axis and semi-minor axis of contours 
is not really consistent with the $k_{\|}\propto k_{\perp}^{1/3}$ relation.
Both the solid curve (5-point) and the dotted curve
(3-point) show a similar scaling.
The average slope is approximately $\sim 0.5$.
Therefore, we can conclude that true anisotropy is similar to or stronger than this.
Note that the dashed curve (2-point) show a milder anisotropy (or, a steeper slope).
This means that multi-point structure functions are indeed better for steep spectrum.
Although structure functions may not be suitable to reveal true anisotropy for EMHD,
it is promising that multi-point structure functions seem to resolve
steeper spectrum better.

\begin{figure*}[h!t]
\includegraphics[width=0.48\textwidth]{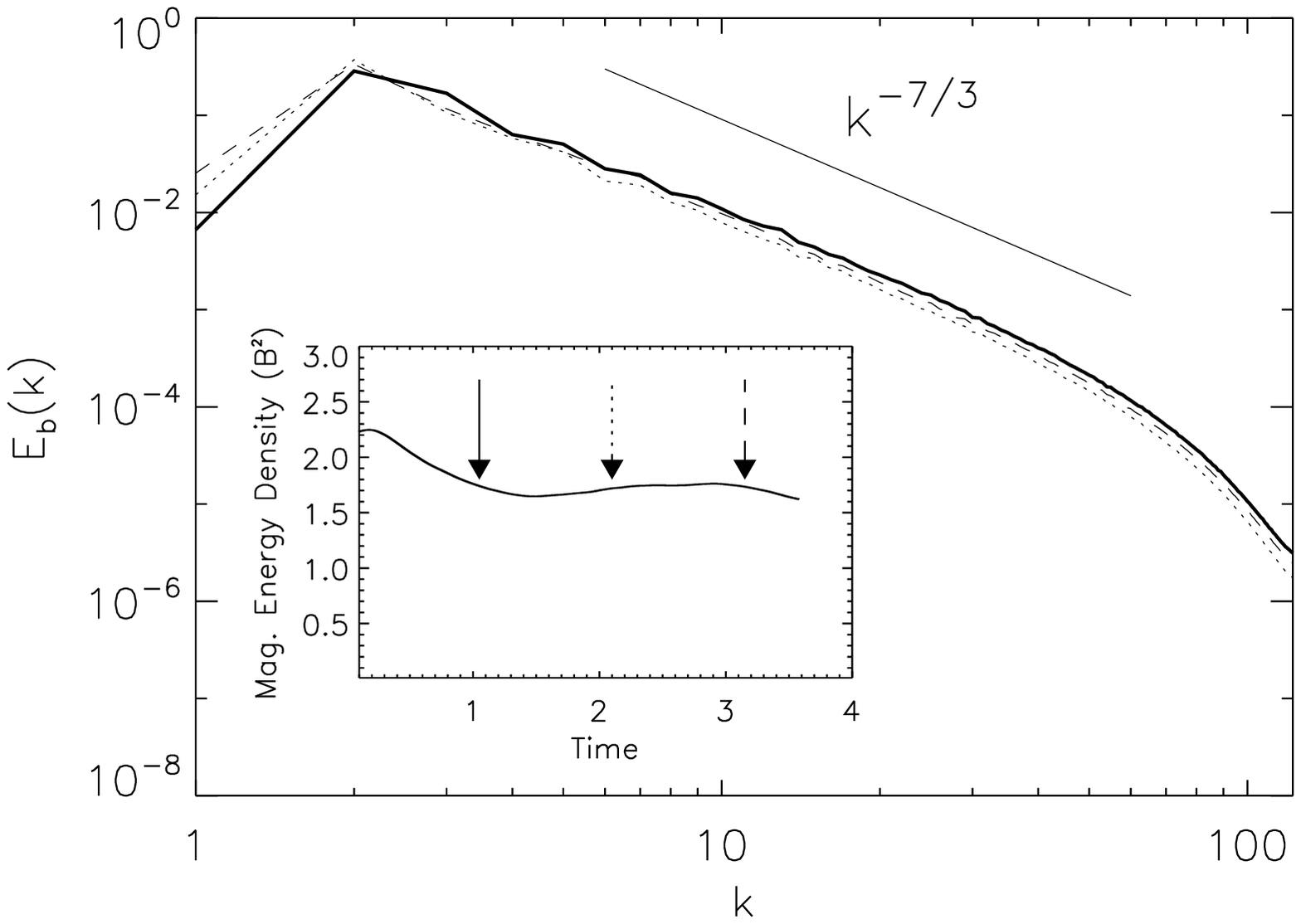}  
\includegraphics[width=0.48\textwidth]{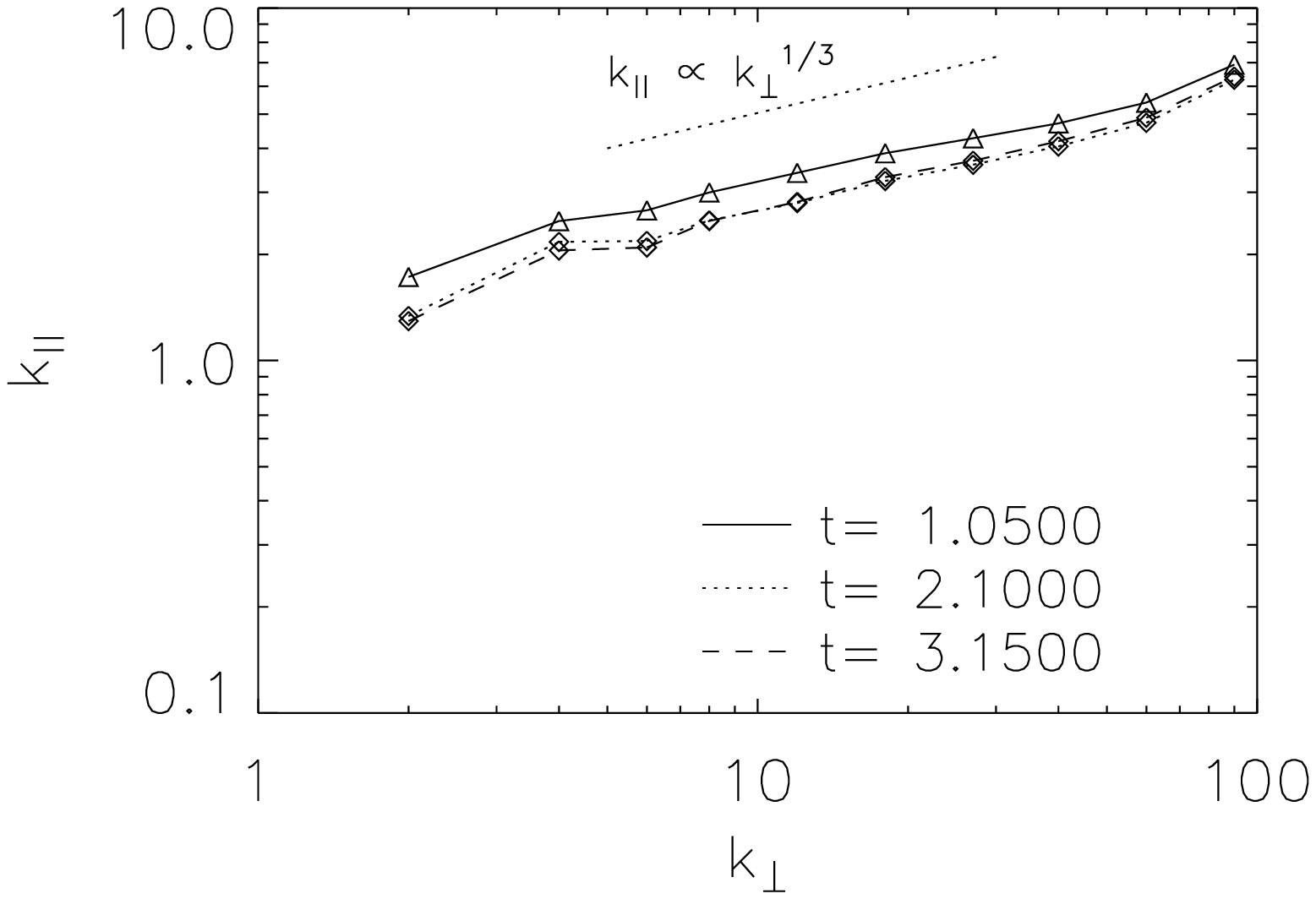}  
\caption{ Driven EMHD turbulence (Run E256F). The strength of
   the (global) mean field is $1$ (i.e.~$B_0=1$).
   The strength of the fluctuating magnetic field ($b$) is maintained to be $\sim 1$.
   Since driving is isotropic, the critical balance is satisfied. 
   {\it Left}: time evolution of magnetic energy density (inset)
   and energy spectra. 
   The spectra are taken at 3 different time points marked by the arrows
   in the inset.
   {\it Right}: anisotropy. Line styles of the curves are the same as those of the arrows
   in the inset.
\label{fig:driven}
}
\end{figure*}

\subsection{Decaying vs. Forced EMHD turbulence}
It is generally true that both decaying and driven turbulence 
show a similar scaling.
However, in the presence of strong mean magnetic field, it is not
clear whether or not they really show a similar scaling.
Consider a decaying strongly magnetized EMHD turbulence.
Let us assume that critical balance is roughly satisfied at t=0.
The strength of random magnetic field drops as time goes on.
But, the strength of the mean magnetic field remains same.
Therefore, critical balance will be soon destroyed. 
Of course, there is a narrow window of time during which critical balance is 
satisfied and we can study critically balanced EMHD turbulence during the time
interval.
Nevertheless, it is worth comparing driven turbulence and decaying turbulence.
In this section, we show that forced EMHD turbulence exhibits similar scaling relations as
decaying EMHD turbulence.

Numerical setup for driven EMHD turbulence is similar to that of decaying
EMHD turbulence. Forcing is done in Fourier space.
The forcing term consists of 21 Fourier components 
with $2\leq k \leq \sqrt{12}$. 
The peak of energy injection is at $k\approx 2.5$.
The forcing is statistically isotropic.  
The strength of the mean magnetic field $B_0$ is 1.
We adjusted the amplitudes of the forcing components so that $b \approx 1$.
Therefore, the driven EMHD turbulence satisfies the condition for critical balance.
This run is designated as Run E256F in Table 1.

Inset in Fig.~\ref{fig:driven} shows time evolution of magnetic energy density 
(in fact $B^2=B_0^2 + b^2$).
The magnetic energy reaches a
stationary state after $t\sim 1$.
The main plot of Fig.~\ref{fig:driven} shows magnetic energy spectrum at 3 different
time points marked by the arrows in the inset.
Line styles of the spectra are the same as those of the arrows.
The slopes of all 3 energy spectra are consistent with $-7/3$.

We plot anisotropy in right panel of Fig.~\ref{fig:driven}.
We use the technique described in Section \S\ref{sect_ani1}.
Anisotropy of driven EMHD turbulence is also consistent with
the $k_{\|}\propto k_{\perp}^{1/3}$ scaling.
All in all, driven MHD turbulence shows similar spectrum and anisotropy as
decaying EMHD turbulence.

\begin{figure*}[h!t]
\includegraphics[width=0.48\textwidth]{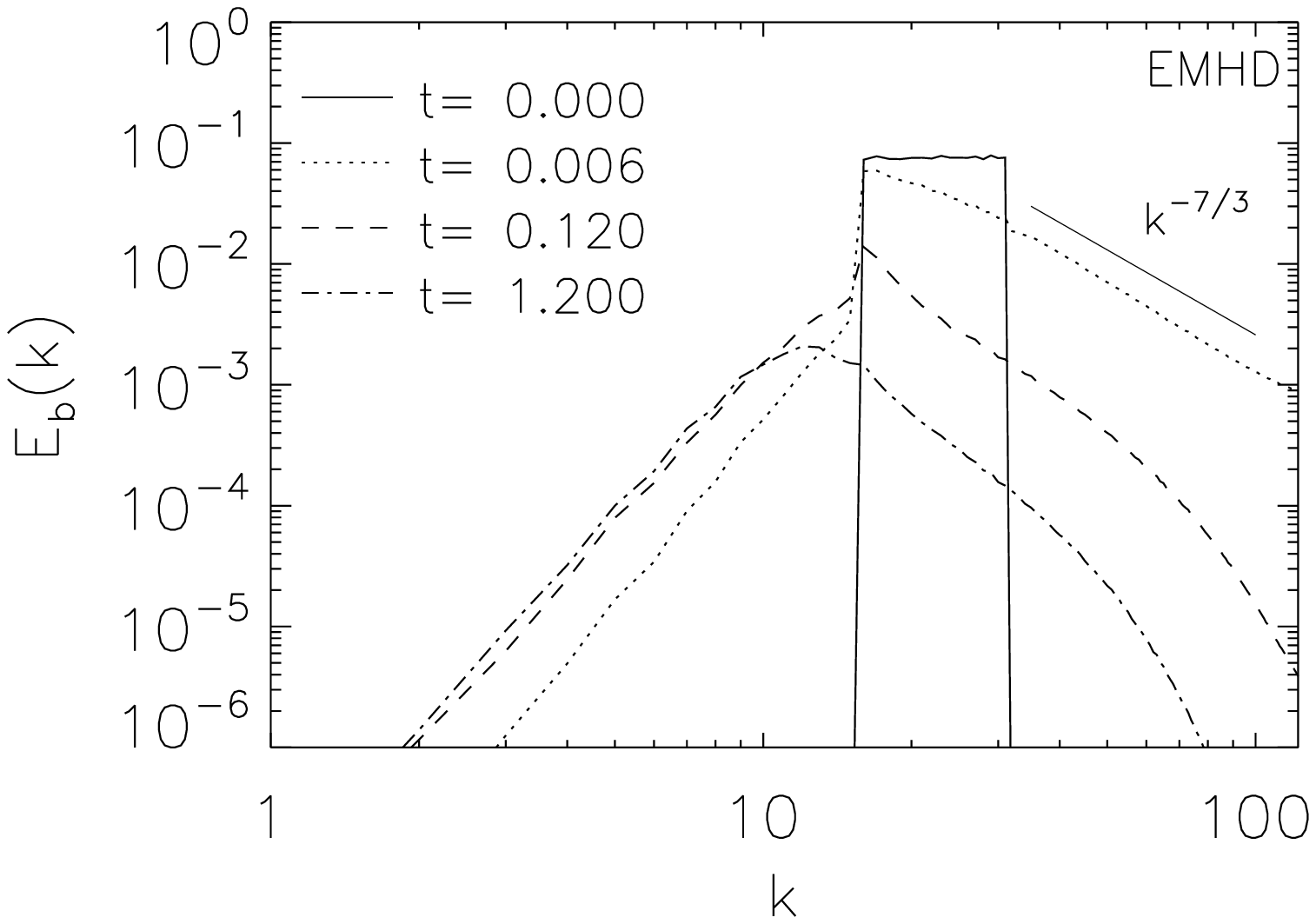}  
\includegraphics[width=0.48\textwidth]{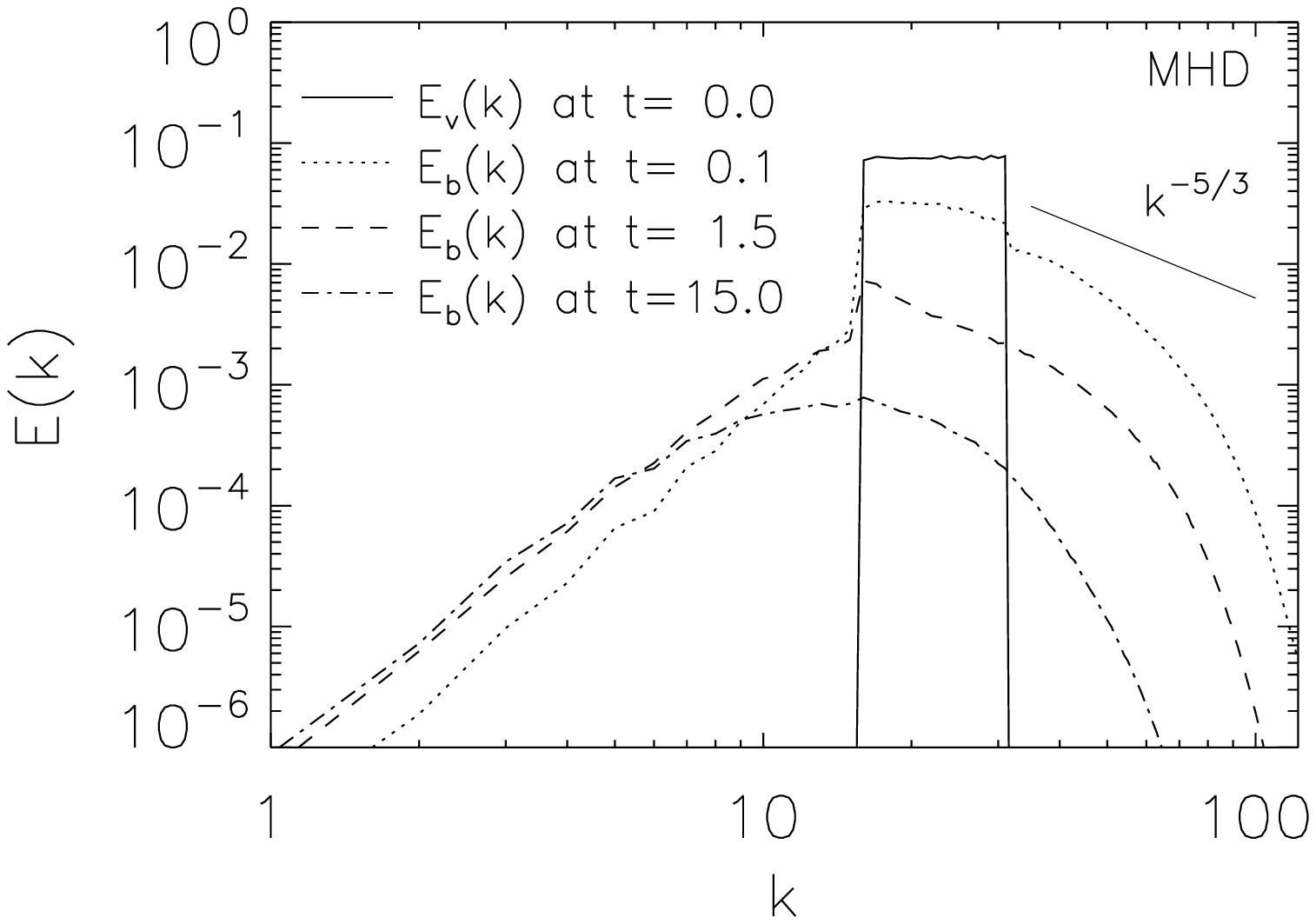}  
\caption{Spectra of decaying turbulence. The mean field ($B_0$) is set to 1 in both cases.
{\it Left}: EMHD turbulence. 
   At $t=0$, the magnetic spectrum is flat for $16\le k \le 32$ (solid curve).
   At later times, energy cascades down to smaller scales.
   At the same time, some energy cascades inversely up to larger scales.
{\it Right}: MHD turbulence. 
   At $t=0$, the {\it velocity} spectrum is flat for $16\le k \le 32$ (solid curve).
   The magnetic field has only the uniform component ($B_0$) at $t=0$.
   At later times, random magnetic field is generated and most of the turbulence 
   energy cascades down to smaller scales.
   We also observe inverse cascade.
   Note that, in the right panel, the solid line is for velocity at t=0 and other lines
   are for magnetic field at later times.
\label{fig:invcas}
}
\end{figure*}

\subsection{Inverse cascade of EMHD turbulence}

In Wareing \& Hollerbach (2009) the inverse cascade
was reported as a part of decaying 2D EMHD turbulence.
The energy spectrum in Shaikh \& Zank (2005) also shows a clear evidence of
inverse cascade in driven 2D EMHD turbulence.
However, the difference of 3D and 2D turbulence
makes one wonder whether the inverse cascade is present in decaying 3D EMHD turbulence.

We perform another decaying EMHD turbulence simulation.
The numerical setup is almost identical to the one described in Section \S4.1.
However, we use a different initial condition.
At $t=0$, Fourier modes with $16\le k \le 32$ are excited (see the solid curve
in Left panel of Fig.~\ref{fig:invcas}).
The numerical resolution is $256^3$.
This run is not listed in Table 1.

Left panel of Fig.~\ref{fig:invcas} shows spectral behavior of the decaying EMHD turbulence.
The dotted line shows the spectrum shortly after the simulation.
The dotted line clearly shows that most of the energy cascades down to small
scales. Of course, the energy cascading down to small scales will dissipate away
below $k\gtrsim 80$.   It also shows that some of the energy 
goes to larger scales, the amount of which is smaller than that in the 2D EMHD case 
(see Wareing \& Hollerbach 2009).
At later times, the peak to the energy spectrum moves to larger scales, which
probably means that we see a self-similar decaying of energy that leads to
the increase of the integral scale (see, for example, Biskamp 2003), rather than
a signature of the inverse energy cascade that is observed in 2D hydrodynamic turbulence.
Nevertheless, we clearly observe that EMHD turbulence can generate 
more coherent magnetic field 
{}from a smaller scale, hence less coherent, magnetic field.
Although this phenomenon is not the inverse cascade per se, 
we can still call it inverse cascade because
it does show that small amount of energy goes to larger scales.
However, in order to avid confusions, we will use the term `small amount of inverse cascade'
whenever possible.

We compare our results with the spectral behavior of the decaying 3D  
MHD turbulence in Right panel of Fig.~\ref{fig:invcas}. 
At $t=0$ velocity field has a flat spectrum between $k=16$ and $k=32$ (solid curve).
The magnetic field at $t=0$ has only the uniform component, the strength of which is set to 1.
The numerical resolution is also $256^3$ and
this run is not listed in Table 1.
At later times, the velocity field generates fluctuating magnetic field mainly 
between $k=16$ and $k=32$.
We see that MHD turbulence also shows small amount of inverse cascade.
Overall spectral behavior of decaying MHD turbulence is similar to that of 
decaying EMHD turbulence.
However, there are also differences.
For example, the gradual shift of the peak of the energy spectrum to larger scale is less
pronounced in the MHD case.
The slopes of the energy spectra on large scales are also different.

\begin{figure*}[h!t]
\includegraphics[width=0.48\textwidth]{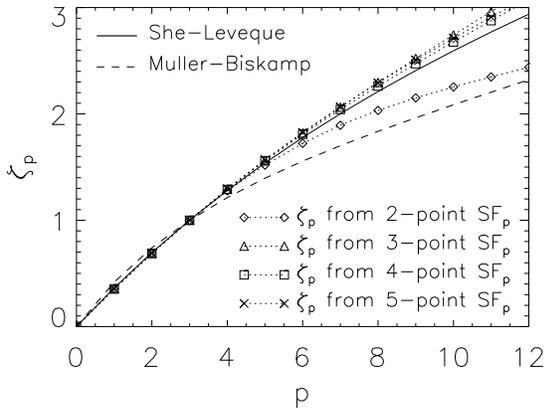}  
\caption{ Intermittency (Run E512D). 
 Scaling exponents from the 3-point, the 4-point, and the 5-point
 structure functions follow the She-Leveque scaling.
 However, those from the standard 2-point structure function show a different scaling.
 The $\zeta_p$'s shown here are the relative scaling exponents, $\zeta_p/\zeta_3$.
\label{fig:zeta}
}
\end{figure*}

\section{High-order Statistics and Bispectrum of EMHD}
In this section, we present other statistical properties of
EMHD turbulence. We use a data cube from Run E512D, which is the
the same data cube as we considered in Section \S\ref{sect_512}.

\subsection{High-order structure functions}
High-order structure functions 
are used for the study of intermittency, which refers to the non-uniform distribution of structures.
The structure functions of order $p$ for magnetic field is defined by
\be
   SF_p(r)=< |{\bf B}({\bf x})-{\bf B}({\bf x}+{\bf r})|^p>_{avg. over~x}.
   \label{eq:sfn}
\ee
Traditionally, researchers use
high-order structure functions of velocity to probe 
dissipation structures of turbulence.
In fully developed hydrodynamic turbulence, the (longitudinal)
velocity structure functions
$SF_p=< ( [ {\bf v}({\bf x}+ {\bf r}) -
      {\bf v}({\bf x})]\cdot \hat{\bf r} )^p>
\equiv < (\delta v_r)^p >$ are
expected to scale as $r^{\zeta_p}$.
One of the key issues in this field is the functional form of the
scaling exponents $\zeta_p$.
There are several models for $\zeta_p$.
Roughly speaking, the dimensionality 
of the dissipation structures plays an important role.

Assuming 1-dimensional worm-like 
dissipation structures, She \& Leveque (1994) 
proposed the scaling relation
\be
  \zeta_p^{SL}=p/9+2[1-(2/3)^{p/3}]
\ee
for incompressible hydrodynamic turbulence.
On the other hand, assuming 2-dimensional sheet-like dissipation structures,
M\"uller \& Biskamp (2000) proposed the relation
\begin{equation}
\zeta_p^{MB}=p/9+1-(1/3)^{p/3}
\end{equation}
for incompressible magneto-hydrodynamic turbulence.
There are other models for intermittency.
But, in this paper, we only consider above mentioned two models
because they are relevant to incompressible variables.

In this paper, we only consider structure functions in perpendicular directions.
That is, in our calculations, the vector ${\bf r}$ (see Eq.~[\ref{eq:sfn}])
is perpendicular to the direction of the local mean field, ${\bf B}_L$.
The reason why we do not consider parallel direction is that
structure functions are not suitable for revealing true scaling relation 
in parallel directions.
In Fig.~\ref{fig:zeta} we plot the relative 
scaling exponent $\zeta_p/\zeta_3$ for EMHD turbulence\footnote{
   In hydrodynamic turbulence, the relative scaling exponents are expressed
   relative to $\zeta_3$ since Kolmogorov's four-fifth law is valid for 
   the third-order longitudinal structure function: $SF_3 \propto r$, where
   the constant of proportionality is (-4/5) times the energy injection rate.
   Therefore, it is natural to use $\zeta_3$ for the relative scaling exponents.
   In MHD, there also exists a similar exact relation 
   for a third-order structure function
   expressed in terms of Elsasser variables
   (Politano \& Pouquet 1998).
   However, according to the exact correlation law obtained by 
   Galtier (2008), there is no simple relation between $SF_3$ and $r$ in EMHD
   and, therefore, the use of $\zeta_3$ for the relative
   scaling exponents is not justified.
   Nevertheless, for the sake of comparison with existing models, 
   we use $\zeta_3$ for the Figure.
}.
It is interesting that the 2-point structure functions
show a different scaling exponents compared with other multi-point structure functions.
This result is not surprising because we already observed that 
the 2-point second-order structure function shows a different behavior 
compared with other multi-point structure functions in right panel of
Fig.~\ref{fig:conto_kpar}.
The scaling exponents based on the 3-point, the 4-point, and the 5-point structure functions
are consistent with the She-Leveque model.

\begin{figure*}[h!t]
\includegraphics[width=0.32\textwidth]{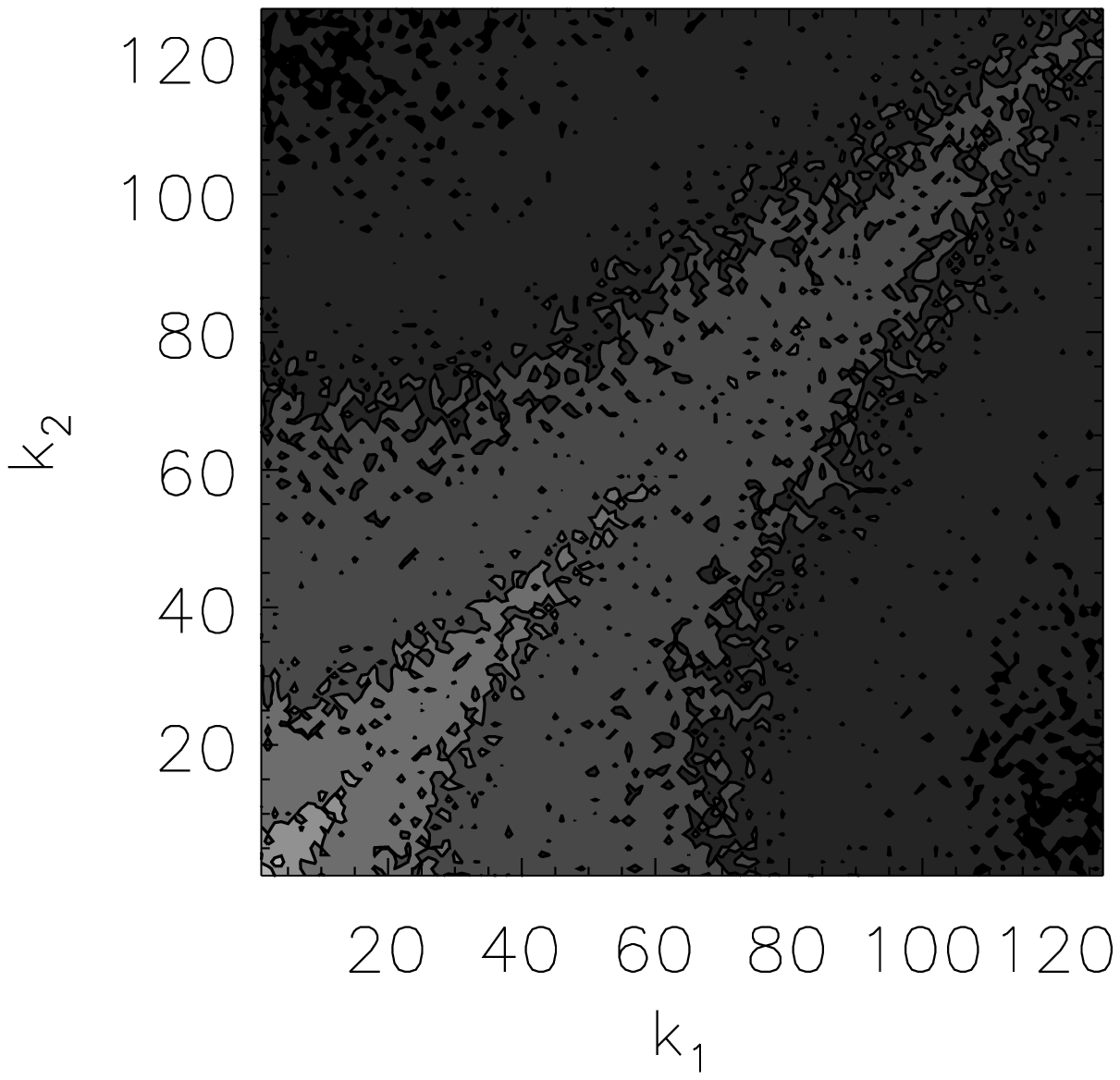}  
\includegraphics[width=0.32\textwidth]{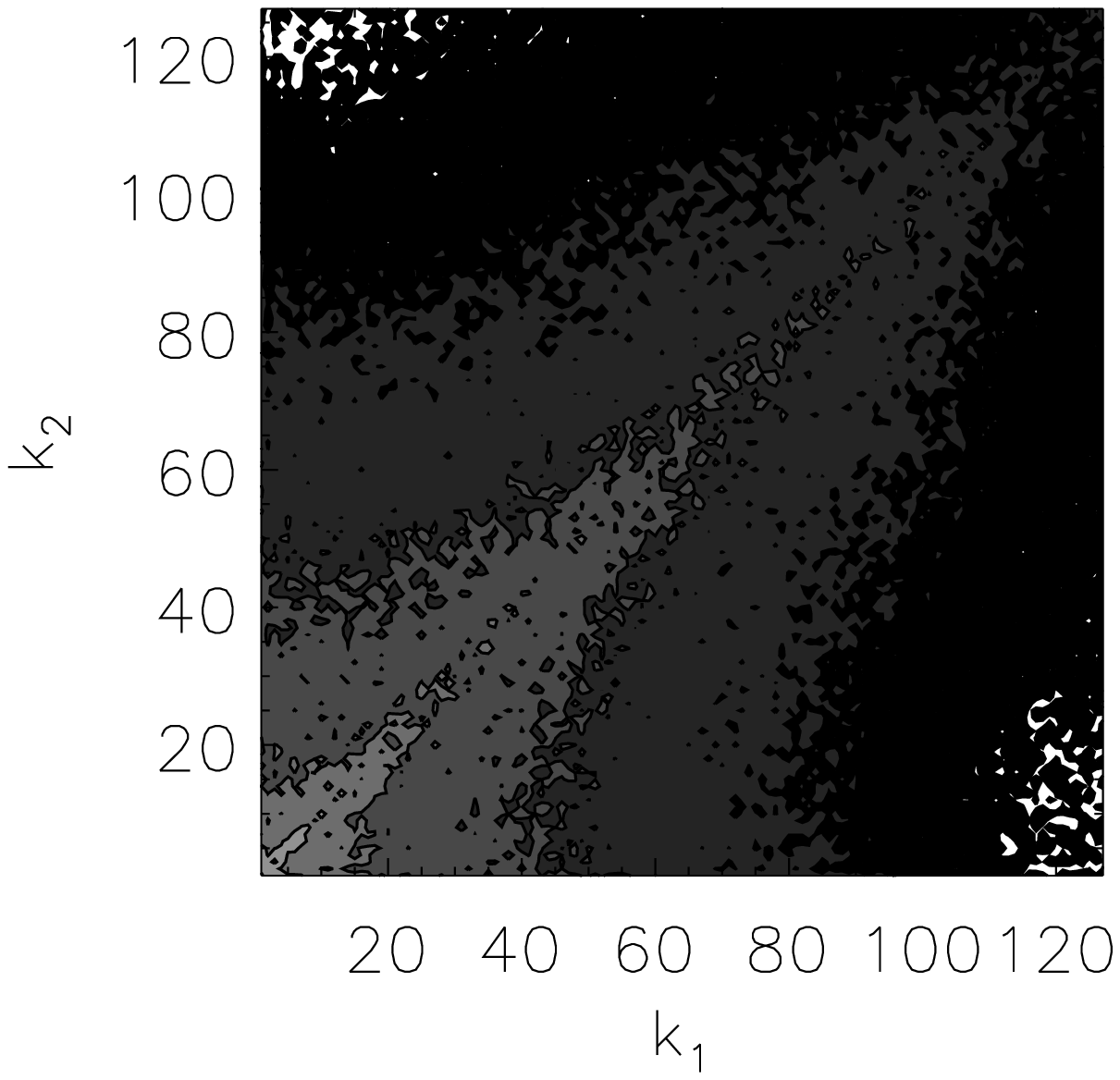}  
\includegraphics[width=0.32\textwidth]{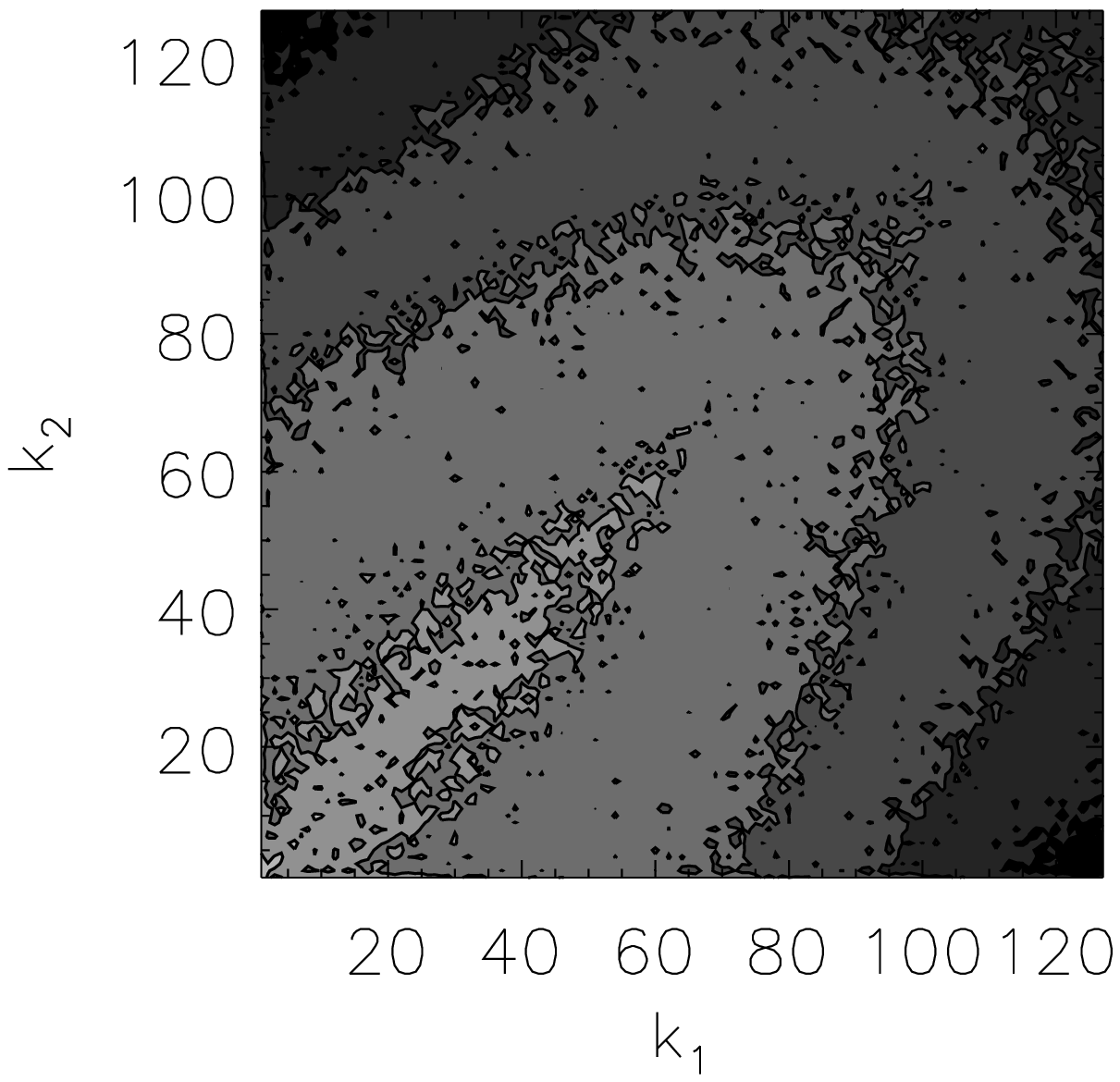}  
\caption{ Bispectrum of EMHD (left and middle panels) and standard MHD (right panel).
  Driven EMHD turbulence at $t\sim 3$ (left panel; Run E256F) and 
  decaying EMHD turbulence at $t\sim 3$ (middle panel; Run E256D) have similar bispectra.
  However, driven MHD turbulence at $t\sim 45$ (right panel; Run MHD256F)   
  shows a different bispectrum.
\label{fig:bisp}
}
\end{figure*}

\subsection{Bispectrum}
In astronomy, the bispectrum is a tool widely used in cosmology and gravitational wave studies
(Fry 1998; Scoccimarro 2000; Liguori et al. 2006).
Here, we briefly describe the definition of the bispectrum (see Burkhart et al. 2009).
The bispectrum is closely related to the power spectrum. 
The Fourier transform of the second-order cumulant, i.e. the autocorrelation function, 
is the power spectrum while the Fourier transform of the third order cumulant 
is known as the bispectrum. In a discrete system, the bispectrum is defined as:
\begin{equation}
BS(\vec{k_{1}},\vec{k_{2}})=\sum_{\vec{k_{1}}=const}\sum_{\vec{k_{2}}=const}\tilde{A}(\vec{k_{1}})\cdot\tilde{A}(\vec{k_{2}})\cdot\tilde{A}^{*}(\vec{k_{1}}+\vec{k_{2}})
\label{eq:bispectra}
\end{equation}
\noindent
where $k_{1}$ and $k_{2}$ are the wave numbers of two interacting waves, and $A(\vec{k})$ is the original discrete data with finite number of elements with $A^{*}(\vec{k})$ representing the complex conjugate of $A(\vec{k})$. As is shown in Equation~(\ref{eq:bispectra}), the bispectrum is a complex quantity which will measure both phase and magnitude information between different wave modes. 

Original formalism for bispectrum is suitable for scalar variables.
Since we deal with magnetic field, which is a vector quantity, we simply
take z-component of magnetic field. Note that ${\bf B}_0=B_0 \hat{\bf z}$.
In Fig.~\ref{fig:bisp}, we plot bispectrum of driven EMHD (left panel),
decaying EMHD (middle panel), and driven MHD (right panel).
For the driven ERMHD and MHD turbulence cases, we take the data after turbulence 
has reached statistically stationary state ($t\sim 3$ and $t\sim 45$, respectively).
The decaying EMHD case was taken at $t\sim 3$ 
(see the inset of Fig.~\ref{fig:e_n_ermhd}).
Both driven EMHD (left panel) and decaying EMHD (middle panel) cases
show similar bispectrum.
However, driven MHD (right panel) shows a substantially different bispectrum.
All cases show that amplitudes of bispectra are highest when $k_1=k_2$, which means 
a high correlation
of modes at $k_1=k_2$.
The amplitudes of bispectra, hence strengths of nonlinear interactions,
drop as we move away from the diagonal lines 
(i.e.~the lines of $k_1=k_2$).
The shape of isocontours depends on the nature of nonlinear interactions.
The case of the standard MHD (right panel) shows wider isocontours than EMHD cases,
which means nonlinear interactions in standard MHD turbulence 
drop more slowly as we move away from the $k_1=k_2$ line.
This demonstrates that nonlinear interactions in ERMHD and MHD cases are
very different.

\begin{figure*}[h!t]
\includegraphics[width=0.7\textwidth]{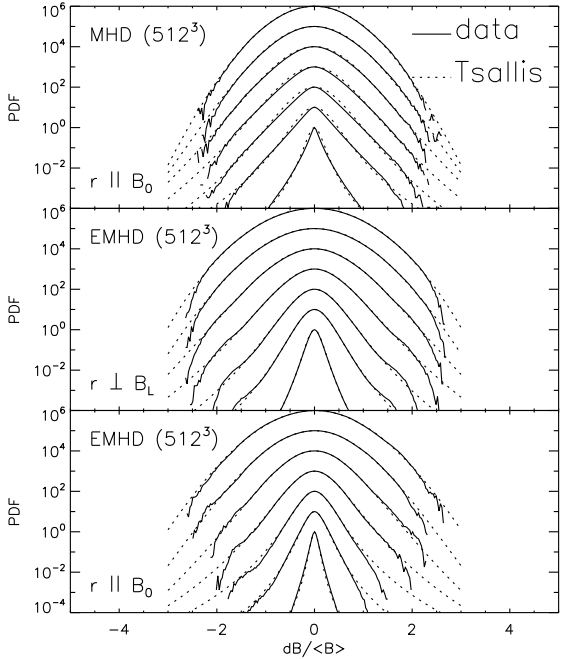}    
\caption{ The PDFs for increments of magnetic field and Tsallis fit.
 The Tsallis distribution fits the PDFs well.
 The PDF is close to a Gaussian distribution when separation $r$ is large (upper curves 
 in each panel) and it deviates from a Gaussian function as the separation gets
 smaller (lower curves).
 Runs E512D (middle and lower panels) and MHD512F (upper panel) are used.
 Note that $B_0$ denotes the global mean field and $B_L$ the local mean field.
 In each panel, curves correspond to  $r=$
  2 (lowermost curve), 6, 10, 17, 29, 50, 88 (uppermost curve), respectively.
\label{fig:pdf}
}
\end{figure*}
\begin{figure*}[h!t]
\includegraphics[width=0.48\textwidth]{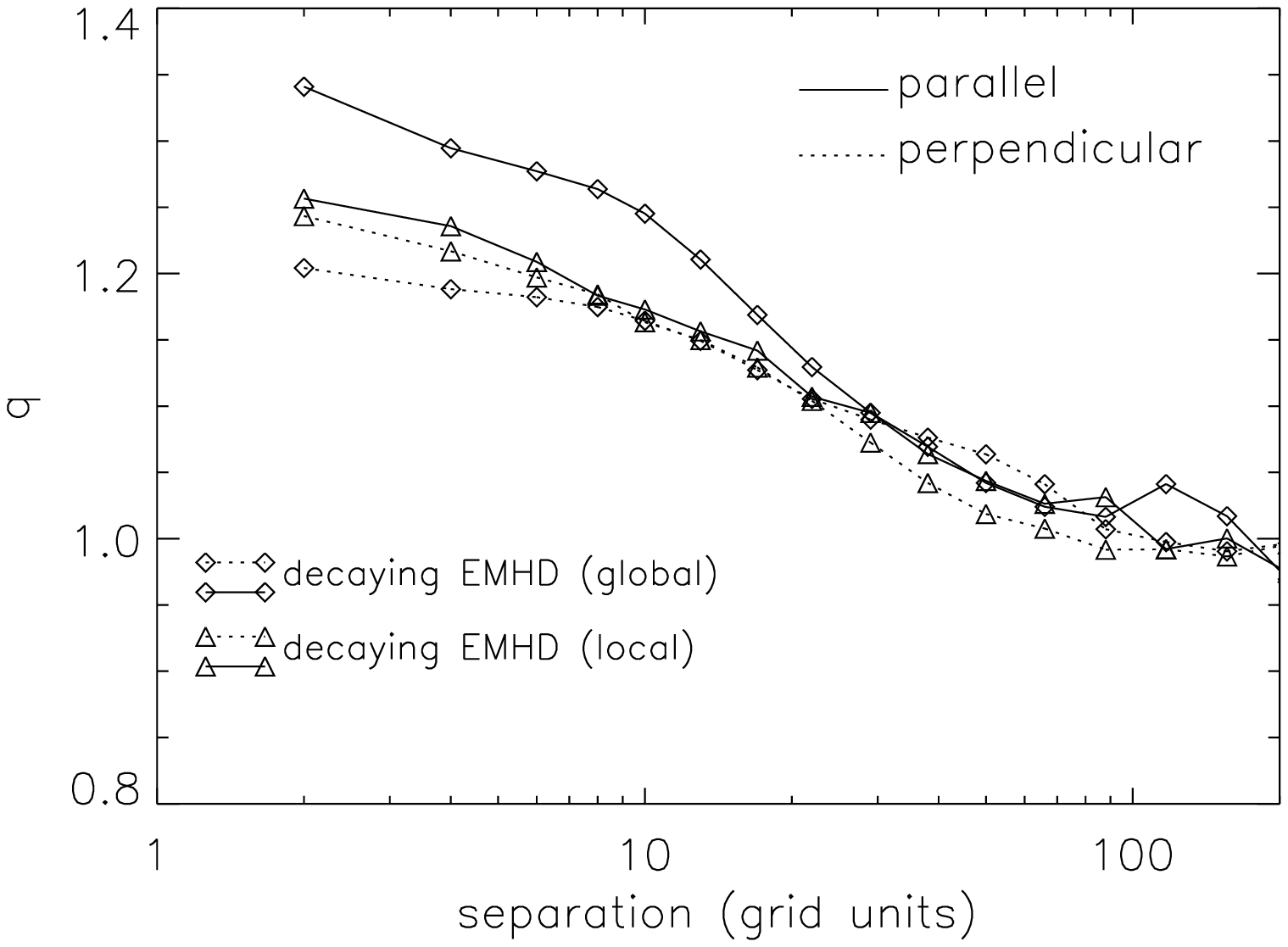}  
\includegraphics[width=0.48\textwidth]{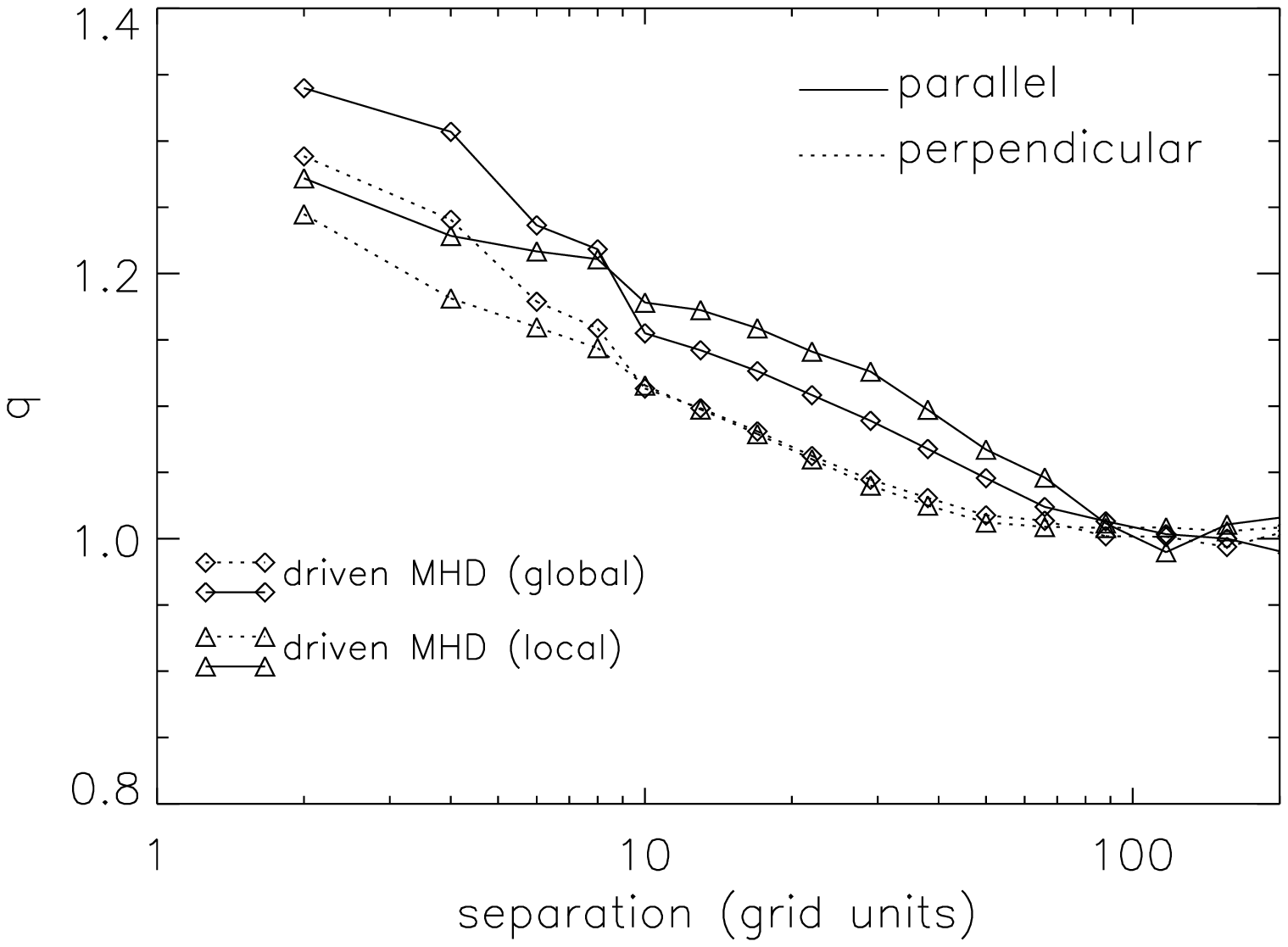}  
\caption{ The $q$ parameter of Tsallis distribution.
  When the separation is large the values of $q$ are very close to $1$, while
  they are larger for smaller separations.
  Note that $q=1$ corresponds to a Gaussian distribution.
  Runs E512D (left panel) and MHD512F (right panel) are used.
\label{fig:q}
}
\end{figure*}
\begin{figure*}[h!t]
\includegraphics[width=0.48\textwidth]{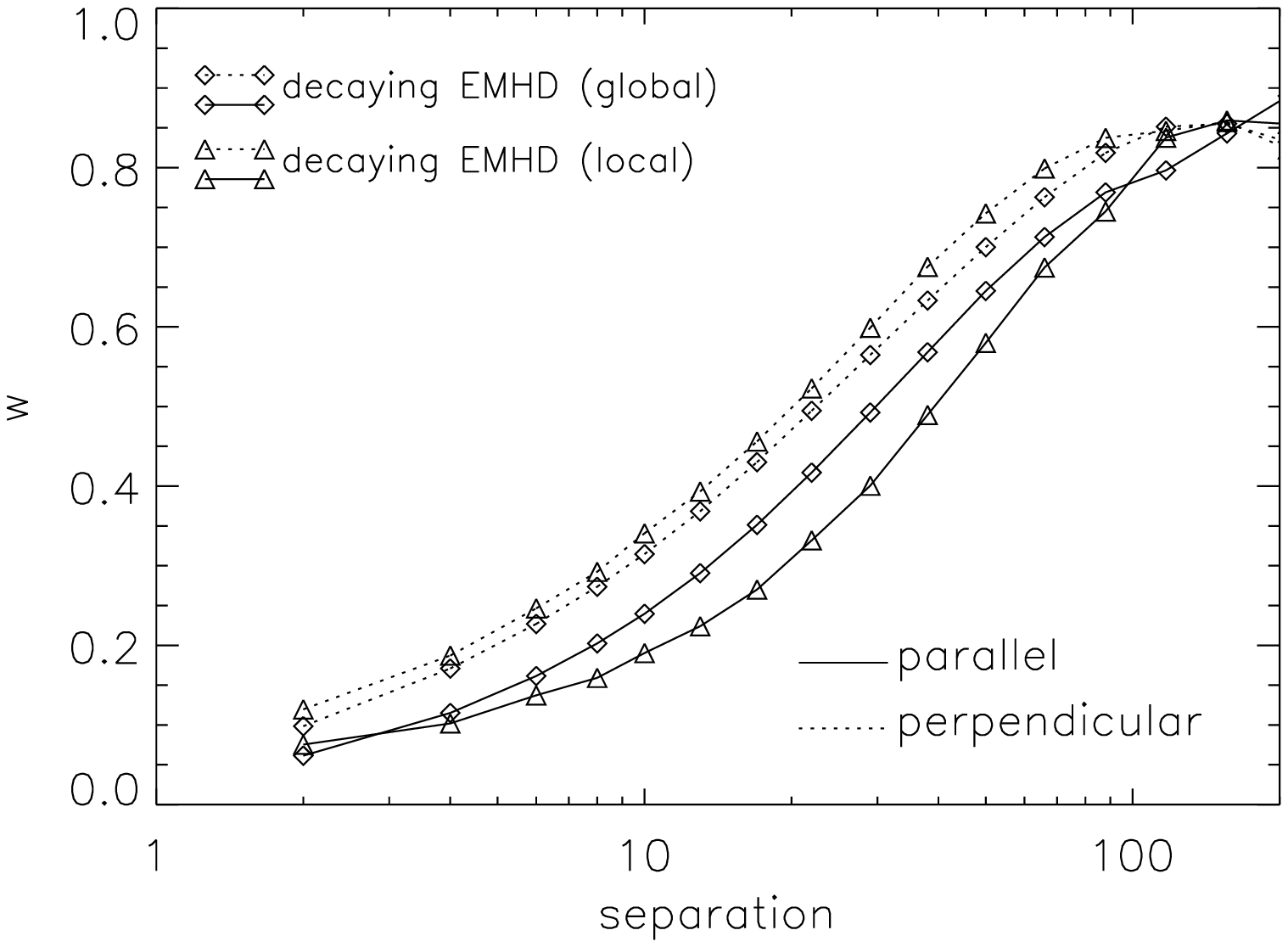}  
\includegraphics[width=0.48\textwidth]{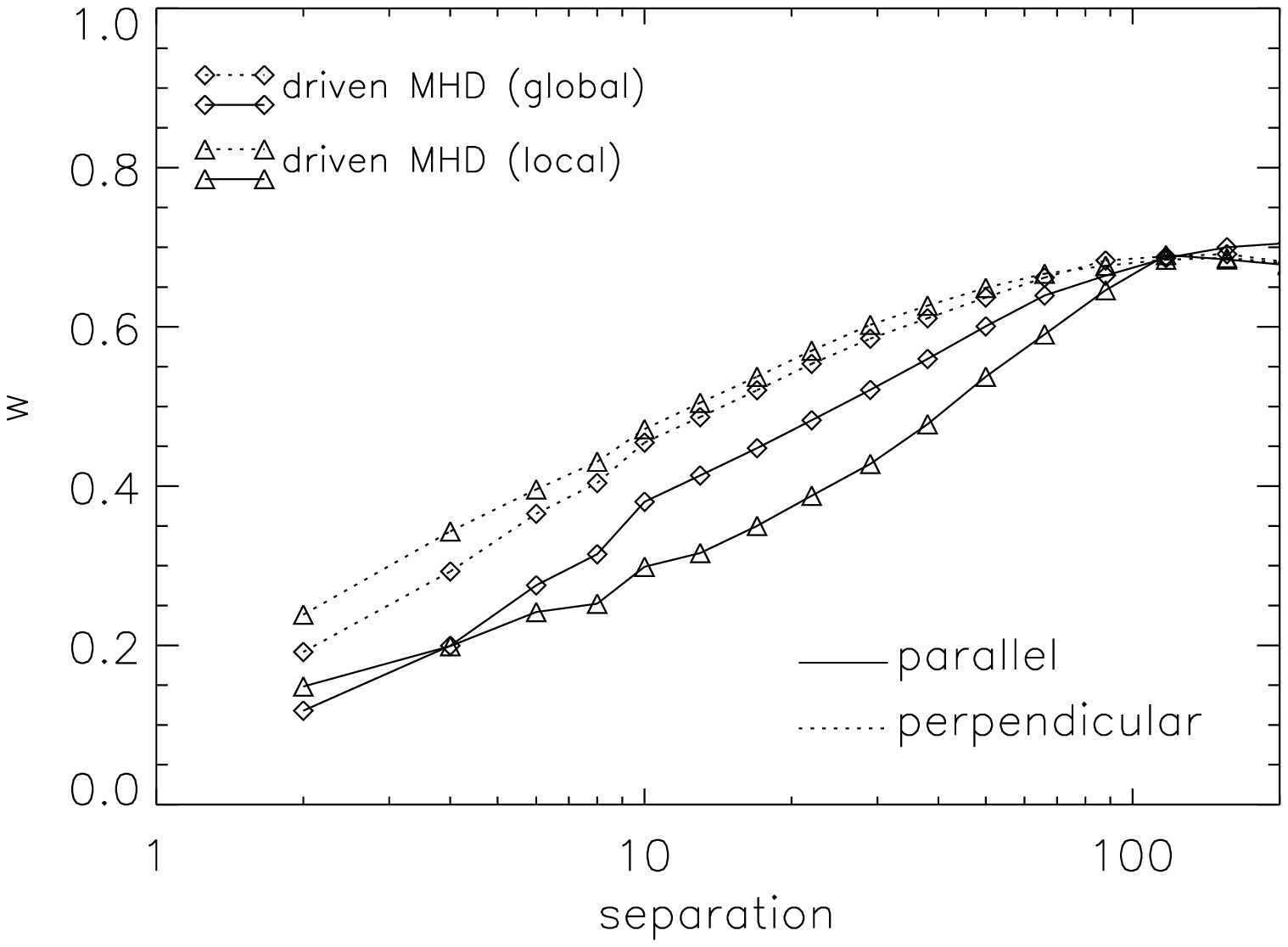}  
\caption{ The $w$ parameter of Tsallis distribution.
  When the separation is larger the values of $w$ is larger.
  The EMHD case (left panel) shows a stronger dependence on the separation.
  Runs E512D (left panel) and MHD512F (right panel) are used.
\label{fig:w}
}
\end{figure*}

\begin{figure*}[h!t]
\includegraphics[width=0.48\textwidth]{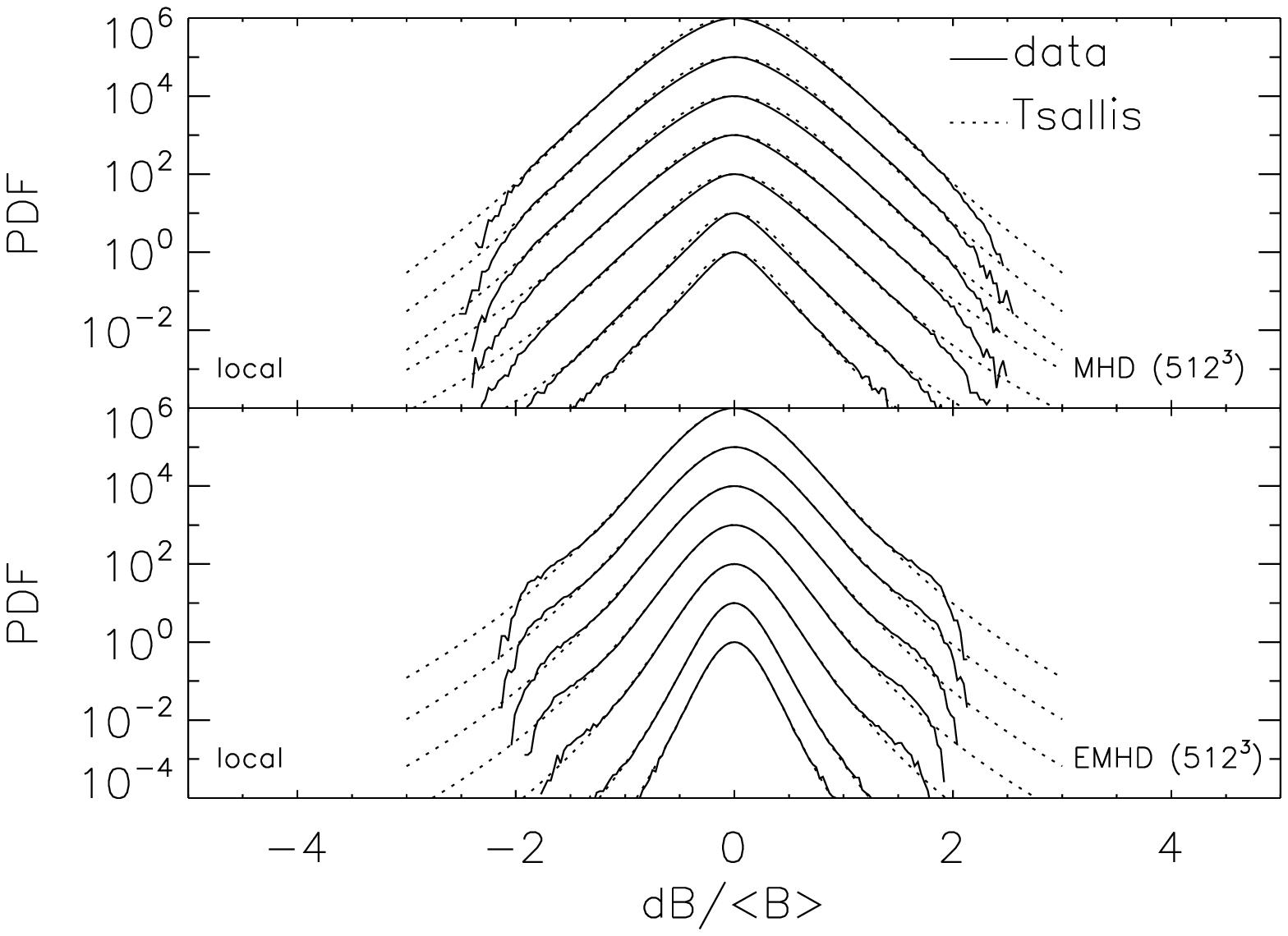}  
\includegraphics[width=0.48\textwidth]{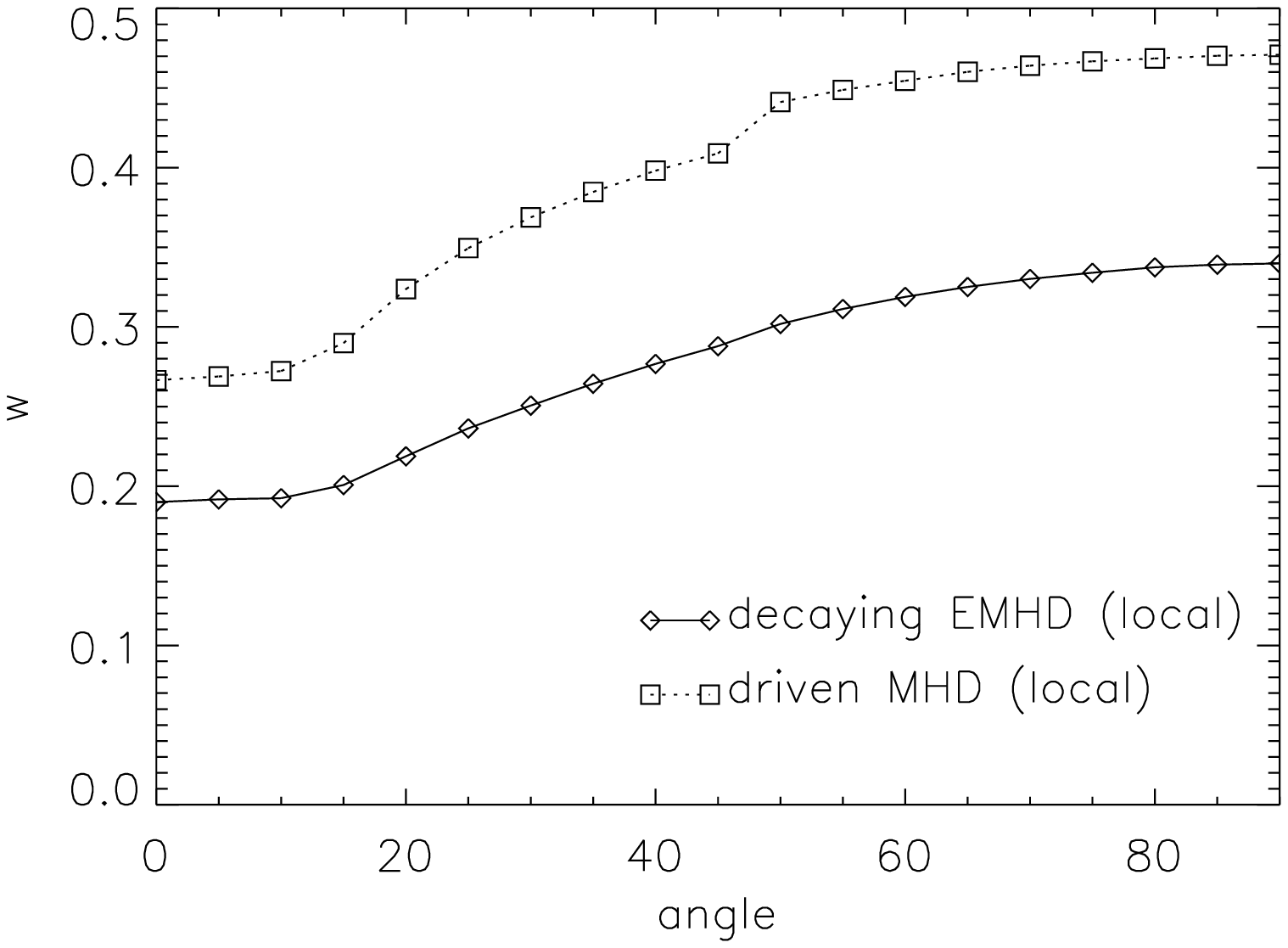}  
\caption{ Angular dependence of the $w$ parameter of Tsallis distribution.
  {\it Left}: The PDFs for $dB$ and Tsallis fit.
  The curves in each panel correspond to $\theta=0^\circ$ (lowermost curve), 
  $15^\circ, 30^\circ, 45^\circ, 60^\circ,
  75^\circ$, and $90^\circ$ (uppermost curve), respectively.
  $\theta$ is the angle between {\it local} mean magnetic field and the direction of ${\bf r}$.
  {Right}: Dependence of $w$ on $\theta$. The horizontal axis is $\theta$ in degrees.
  Runs E512D (EMHD) and MHD512F (MHD) are used. 
  All calculations are done in local frame.
\label{fig:angle}
}
\end{figure*}

\section{Comparison with the solar wind turbulence}
In this subsection, we compare statistics of MHD/EMHD turbulence and
turbulence in the solar wind.
We consider quantities related to the probability distribution functions (PDFs)
of fluctuations in increments of the magnetic field strength $B$
\be
   dB\equiv \langle B({\bf x}+{\bf r})-B({\bf x})\rangle_x /\langle B({\bf x}) \rangle_x,
   \label{eq:db}
\ee
where $B=|{\bf B}|$ is the strength of magnetic field and 
the angled brackets $\langle...\rangle_x$ denote average taken over ${\bf x}$.
In this section, we are only concerned with characteristics of turbulence
directly measured in space plasmas.

\subsection{PDFs for dB}
In the solar wind, the PDFs for increments of the magnetic field strength, $dB$,
 is measured by
\be
   dB(\tau) \equiv \langle B(t+\tau)-B(t)\rangle_t /\langle B(t) \rangle_t,
\ee
where $B(t)$ is the strength (or the averaged strength) of magnetic field  
at time $t$ and $\langle...\rangle_t$ denotes average taken over time.
The observed PDFs for $dB$
on scales from 1 hour to 128 days 
are well described by the Tsallis distribution
(Burlaga, F-Vinas, \& Wang 2007; Burlaga \& F-Vinas 2005), which
is given by
\be
   y(x)=A\left[ 1+(q-1)\frac{ x^2 }{ w^2 } \right]^{-1/(q-1)},
\ee
where $A$, $q$, and $w$ are constants. 
The function is proportional to a Gaussian function for small $x$,
\be
  y(x) \approx A\left( 1-\frac{x^2}{w^2}\right) \approx A\exp(-\frac{x^2}{w^2})
  \mbox{,~~as $x\rightarrow 0$},
\ee
and a power law for large $x$,
\be
  y(x) \propto x^{-2/(q-1)} \mbox{,~~as $x\rightarrow \infty $}.
\ee
The shape of the function is determined by the values of $q$ and $w$.
In the limit of $q \rightarrow 1$, the function reduces to a Gaussian distribution.
Therefore the value of $q$ is a measure of non-Gaussianity.
The value of $w$ is related to the width of the distribution.
Since $\tau > 1$ hour, the measured PDFs in Burlaga et al. (2007) and
Burlaga \& F-Vinas (2005) reflect fluctuations of MHD turbulence.

Fig.~\ref{fig:pdf} shows PDFs of MHD (top panel) and EMHD (middle and bottom panels)
turbulence.
The solid lines are PDFs of $dB$ in our numerical data
and the dotted curves are fits of the PDFs to the Tsallis distribution.
We use the Levenberg-Marquardt algorithm 
(Levenberg 1944; Marquardt 1963; see Press et al.~1992) for fitting.
Each curve in the panels represents the PDF for a particular separation.
{}As we move from the lowest curve in each panel, the separations in grid units are
2, 6, 10, 17, 29, 50, 88, respectively.
Each curve is displaced vertically from the one below by a factor of 10.
We set the value of each PDF at $dB=0$ to 1 before displacement, for
the sake of clarity. 

The PDFs may depend on the direction of ${\bf r}$ in Eq.~(\ref{eq:db}).
That is, the PDFs for the parallel direction can be different from those for
perpendicular direction.
Therefore, we calculate PDFs for parallel and perpendicular directions separately.
We try both global and local frames to define parallel and perpendicular directions. 
Overall, we consider 4 cases: parallel direction in global frame ($r || B_0$),
parallel direction in local frame ($r || B_L$),
perpendicular direction in global frame ($r \perp B_0$), and
perpendicular direction in local frame ($r \perp B_L$).

Our calculations show that there is no big difference in the PDFs of the above mentioned
4 cases. General trend is that the PDF is close to a Gaussian function when separation is large
(upper curves in each panel) and it deviates from a Gaussian function when the separation is
small (lower curves in each panel). We note however that, for a given separation (for example,
see the lowest curve in each panel, which corresponds to $r=2$ grid units),
the widths of the PDFs vary, depending on the direction of ${\bf r}$.

We can confirm quantitatively  
the trend that the PDF is close to a Gaussian distribution when separation is large
and it deviates from a Gaussian distribution when the separation is small.
In Fig.~\ref{fig:q}, we plot the values of $q$.
The left panel is for EMHD turbulence (Run E512D) and the right panel for
MHD turbulence (Run MHD512F).
As we mentioned, the value of $q$ is close to $1$ when 
the PDF is close to a Gaussian distribution. 
Indeed, when the separation is large, $q$ is very close to $1$.
The value of $q$ deviates from $1$ as the separation gets smaller, which means that
the PDF deviates from a Gaussian distribution.
The overall trend is consistent with the observations of the MHD-scale fluctuations
in the solar wind
(see Burlaga \& F-Vinas 2005; Burlaga et al. 2007).
But, this should be taken as a very approximate statement, because
 the behavior of $q$ in the solar wind is very complicated.
In each panel of Fig.~\ref{fig:q}, we plot the $q$ values of 4 different directions mentioned above.
In general, all 4 cases show a similar trend, although the $q$ values for the parallel
cases are slightly larger than those for perpendicular cases.

In Fig.~\ref{fig:w}, we plot the values of $w$.
The left panel is for EMHD turbulence (Run E512D) and the right panel for
MHD turbulence (Run MHD512F).
It is interesting that the values of $w$ in EMHD cases show steeper dependence on
separation than those of MHD cases.
One should also note that the behavior of $q$ as well as $w$ for parallel directions in EMHD cases 
may be dominated by
large-scale fluctuations, because the energy spectrum $E(k_{\|})$ is too steep
(see Section \S\ref{sect:sf2} and Appendix B).

Fig.~\ref{fig:w} shows that the values of $w$ for perpendicular direction (dotted lines)
are systematically larger than those for parallel direction (solid lines).
Fig.~\ref{fig:angle} shows that $w$ parameter indeed depends on the direction of ${\bf r}$.
The curves in the left panel are the PDFs for different values of $\theta$, which is the angle
between the {\it local} mean magnetic field and the separation vector ${\bf r}$.
{}The curves in top or bottom panel
correspond to 
  $\theta=0^\circ$ (lowermost curve), 
  $15^\circ, 30^\circ, 45^\circ, 60^\circ,
  75^\circ$, and $90^\circ$ (uppermost curve), respectively.
We can clearly see that the width of the PDF gets larger as the angle increases.
Note that the $w$ parameter of Tsallis distribution is related to the width of the PDF.
The right panel quantitatively shows dependence of $w$ on $\theta$.
The $w$ parameter is smallest when $\theta=0$ (=parallel direction) and it is largest
when $\theta=90^\circ$ (perpendicular direction).

\begin{figure*}[h!t]
\includegraphics[width=0.48\textwidth]{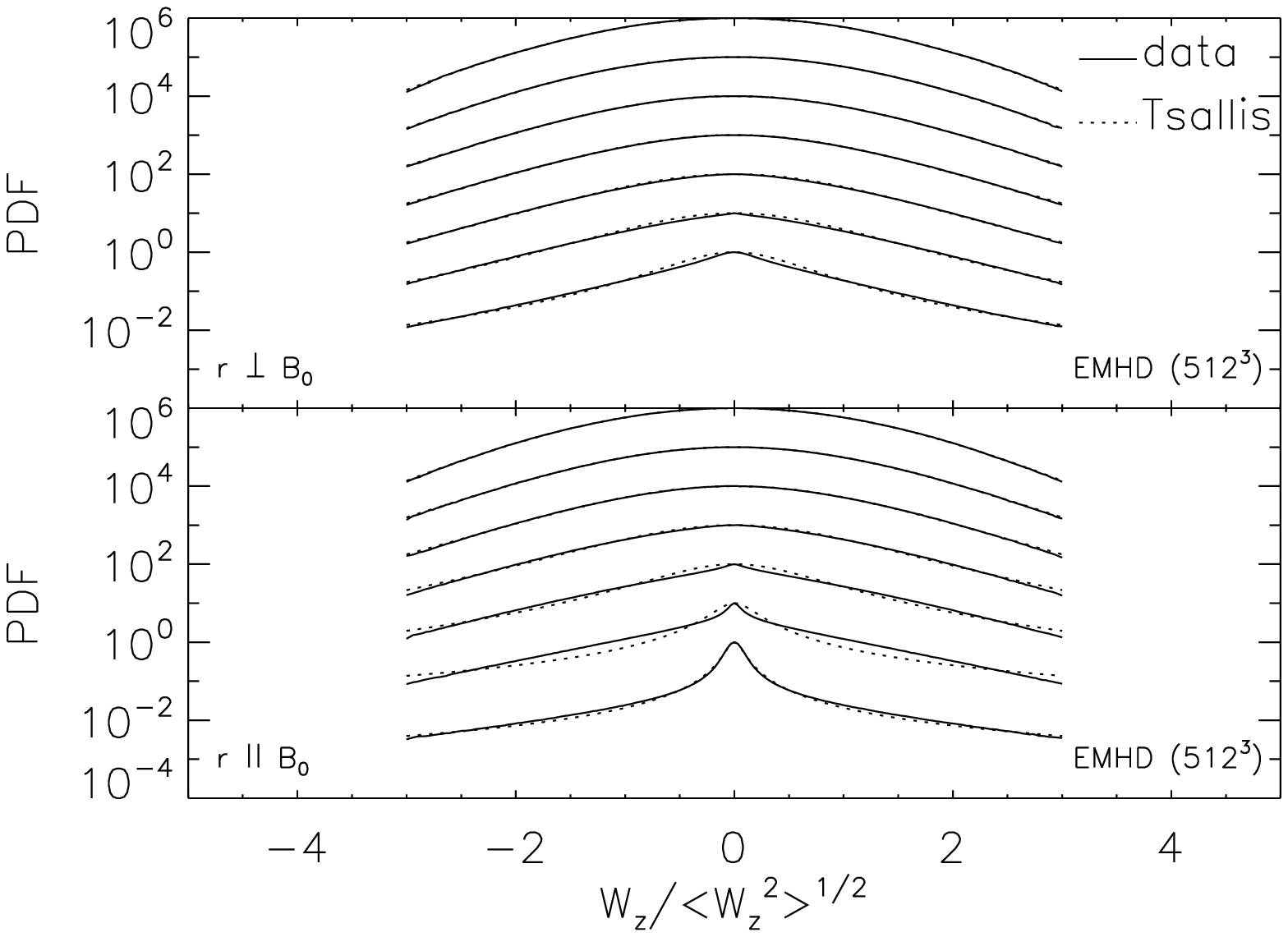}  
\includegraphics[width=0.48\textwidth]{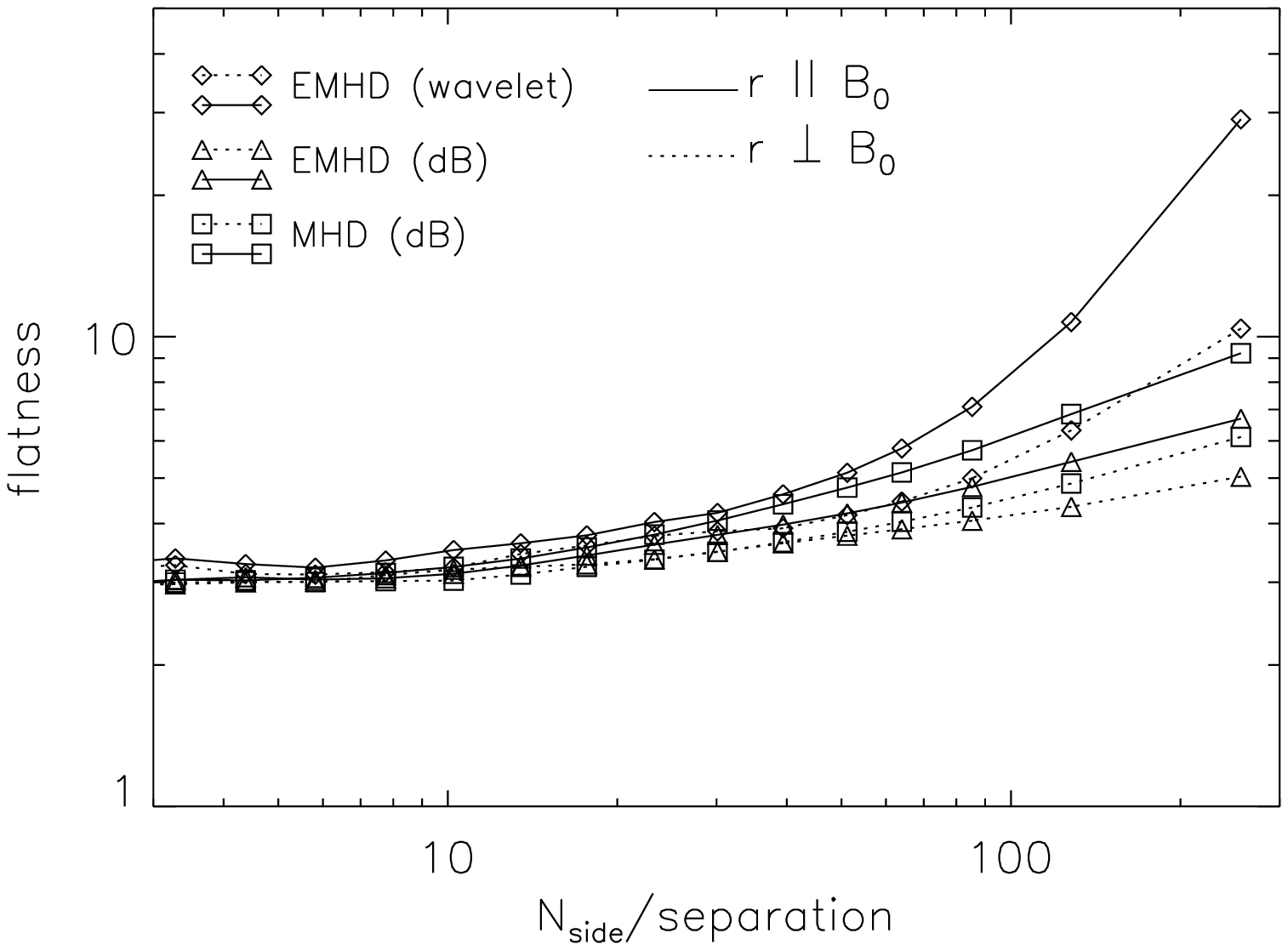}  
\caption{ The PDF and flatness (Kurtosis) of the Morlet wavelet transformation coefficient $\mathcal{W}_z$,
where $z$ is parallel to the global mean field ${\bf B}_0$.
  {\it Left}: the Tsallis distribution approximately fits the PDFs.
 In each panel, curves correspond to  $r=$
  2 (lowermost curve), 6, 10, 17, 29, 50, 88 (uppermost curve), respectively.
  {\it Right}: flatness is larger for smaller separation.
  When the separations are large, the measured values of the flatness are very close to 3,
  which is the same as the flatness of the normal distribution.
\label{fig:wavelet}
}
\end{figure*}

\subsection{PDFs for wavelet transform coefficients}
The wavelet transform is sometimes used for the study of the magnetic field
fluctuation in the solar wind.
For example, Alexandrova et al. (2008) used
the Morlet wavelet transform
\be
   \mathcal{W}_i(\tau,t)=\sum_{j=0}^{N-1} B_i(t_j)\psi \left[ (t_j-t)/\tau \right],
   \label{eq:morlet}
\ee
where $B_i(t_j)$ is the $i$th component of the magnetic field at $t_j=t_0+j\times \Delta t$,
and $\psi(u)\propto \cos(6u)\exp(-u^2/2)$ is the Morlet wavelet.
In this subsection, we briefly consider PDFs of the Morlet wavelet transform 
coefficient $\mathcal{W}_i$.

We perform the Morlet wavelet transform using our numerical data.
We only consider the parallel (or $z$) component of magnetic field in global frame.
Note that, in our numerical calculations, $\tau$ in Eq.~(\ref{eq:morlet}) is  proportional
to separation $r$ and the global mean magnetic field ${\bf B}_0$ is parallel to $z$-direction.
Fig.~\ref{fig:wavelet} shows that the Tsallis distributions roughly fit the PDFs of EMHD
turbulence (Run E512D).
The horizontal axis is $\mathcal{W}_z(\tau)/\langle \mathcal{W}_z(\tau)^2 \rangle_t^{1/2}$.
In general the PDFs for large separations (upper curves in each panel) are close to
Gaussian distributions and those for small separations (lower curves in each panel)
show departure from Gaussian ones.
The PDFs for the parallel direction (lower panel) show stronger departure from 
Gaussian distributions, which is confirmed by the large values of $q$ for small separations
(not shown in this paper).
In the right panel of the Figure, we show the flatness (or Kurtosis) as a function of the scale, which
is defined by
\be
   \langle \mathcal{W}_i^4 \rangle/ \langle \mathcal{W}_i^2 \rangle^2.
\ee
We show both MHD and EMHD cases in the same plot.
In the case EMHD, we show the flatness of both $dB_z$ and $\mathcal{W}_z$
In the case of MHD, we show the flatness of $dB_z$ only.
When the separation is large, the flatness is very close to the Gaussian value of $3$. 
In general, it increases as the separation decreases, which is consistent with
observations (Carbone et al. 2004; Alexandrova et al. 2008; see also Burlaga \& F-Vinas 2005).
Note that flatness of $\mathcal{W}_z$ of EMHD turbulence in the parallel direction 
shows a strong dependence on the separation.

\section{Discussions}

\subsection{Anisotropy of EMHD}

Anisotropy of EMHD is impossible to measure using the traditional second order structure function.
In CL04,
we proposed to a different technique which allowed us to actually get 
$k_{\|}\sim k_{\bot}^{1/3}$, which confirmed our theoretical prediction.

However, as the problem of measuring anisotropy is important, in the present paper we provided not only higher resolution simulations, but also a few new techniques of measuring anisotropy. 
All these techniques provided results consistent with our earlier finding. 

Unlike CL04, in this paper we did not limit ourselves by the simulations of the decaying turbulence, but also studied driven turbulence. We did not observe differences in the slope or spectrum between the two cases.  

\subsection{EMHD and Kinetic Alfven wave turbulence}

As we have shown in \S2, there are two approaches to describing the turbulence over scales less  than the proton gyroscale. The traditional approach assumes that the term proportional to $\nabla\cdot{\bf v}$ (see
Eq.~[\ref{eq:divv}]) is negligible, while the approach in Schekochihin et al. (2009) insists on keeping this term. We performed numerical simulations with and without the term and obtained 
virtually identical results for the turbulence spectrum and anisotropy. 
This result shows importance of EMHD to analyze collisionless plasmas.

CL04 derived the anisotropy of the EMHD turbulence and confirmed the obtained scaling numerically.
Recently, Howes et al. (2008a) used a gyrokinetic code and obtained the expected $k^{-7/3}$ energy spectrum 
for magnetic fluctuations below the proton gyroscale (see also discussions in Matthaeus et al. 2008 and
Howes et al. 2008b). Anisotropy has not been studied with a gyrokinetic code yet.

\subsection{MHD and EMHD turbulence}

Within the paper we have employed a set of different statistical measures to characterize EMHD turbulence. We compared the results with those obtained for MHD turbulence. 
We found both similarities and differences between the two types of turbulence.

MHD turbulence and EMHD turbulence have different energy spectra and anisotropy in spite of the fact
that their scaling relations are derived from the same principles: constancy
of energy cascade and critical balance. 
Our confirmation of the anisotropy scaling predicted in CL04 testifies that the importance 
of the critical balance goes beyond MHD.

Surprisingly, the high-order statistics shows a similar scaling: 
scaling exponents of both MHD turbulence (see Cho, Lazarian \& Vishniac 2002) 
and EMHD turbulence are compatible with 
those of the She-Leveque scaling\footnote{
   Note however that we considered turbulence in strongly magnetized medium and 
   that the $SF_p$ we measured are for directions perpendicular to the local mean field ${\bf B}_L$.
}.
Does it mean that the She-Leveque scaling is so universal that it
fails to distinguish different types of turbulence? What is the underlying reason for this universality? These are still open questions to be addressed by the future research.

Interestingly enough, the PDFs for increments of magnetic field are well described by the Tsallis function. 
The correspondence between the space measurements and Tsallis description can be found, for example, 
in Burlaga \& F.-Vinas (2005),
Leubner \& V\"{o}r\"{o}s (2005), and Burlaga et al. (2007).
Now we confirmed the correspondence with direct numerical simulations. 

Another statistics, namely, the bispectrum is very different for MHD and EMHD. 
The bispectrum measures non-linear interactions. The difference observed confirms 
the difference in cascading of the two types of turbulence.

\subsection{Inverse cascade in EMHD}

Our simulations in \S6.5 show the existence of small amount of 
inverse cascade in decaying EMHD turbulence. 
Although we
have not performed simulations with driven turbulence, 
we expect that small amount of inverse cascade is an 
intrinsic part of the EMHD cascade.
That is, we expect that generation of small amount of
larger scale, hence more coherent, magnetic field from small scale turbulence is
an intrinsic part of the EMHD turbulence.

This kind of inverse cascade has important implications in the 3D MHD case.
When there exists turbulent velocity cascade,
this kind of inverse cascade
enables generation of
coherent field on the outer scale of turbulence 
{}from a small-scale seed magnetic field
(see Cho et al. 2009).
Note, however, that the existence of
turbulent velocity cascade that can amplify the magnetic field is assumed here. 
In EMHD, on the contrary, the fluctuations of magnetic field and velocity are 
connected with each other from the very beginning. 
Nevertheless, one can speculate that the small amount of inverse cascade in EMHD turbulence can 
help to create large scale magnetic structures when the driving is at small scales. 
An interesting implication of this kind of inverse EMHD cascade could be its contribution 
to the creation of the dipole field, when magnetic field is generated by the 
thermomagnetic instability in the crust of a neutron star.

\subsection{Comparison with observations}

Space plasma measurements allow for a direct comparison with the results of turbulence simulations. 
This comparison is essential, as numerically turbulence can be studied only for relatively small Reynolds and Lundquist numbers. As a result, validation of the obtained scaling in realistic astrophysical environments is absolutely essential (see an extended discussion of the issue in Lazarian et al. 2009).

It is encouraging that the spectra and the PDFs of the increments of magnetic field strength
obtained numerically show correspondence with observations. We believe that more detailed comparisons are necessary.

\subsection{EMHD turbulence anisotropy and the ADAF model}
The issue of why accretion disks around black holes are not as luminous as one would expect is a burning
question, addressing of which is required for explaining the low luminosity of black hole environments in the centers of galaxies. One of the ideas proposed was that of Advection Dominated Accretion Flows (ADAFs) 
(Narayan \& Yi 1995; Narayan et al. 1998). 
According to this idea the low luminosity of the accreting material is due to the low rate of the transfer of energy of the turbulent accreting flow to electrons. It is postulated in the model that protons carry the lion's share of the flow energy and thus the emission is suppressed. Is it so?

Quataert \& Gruzinov (1999) discussed transition from
standard Alfvenic MHD turbulence to EMHD turbulence in advection dominated 
accretion flows.
As Alfvenic turbulence reaches the proton gyroradius scale,
some of the turbulence energy goes to protons through collisionless damping.
The major heating mechanism for protons right above the proton gyroscale is
the transit time damping (TTD) caused by non-zero parallel magnetic field
fluctuations.
Heating of protons is sensitive to $\beta_i \approx (v_i/v_A)^2$, 
where $v_A$ is the Alfven speed.
When $\beta_i \gg 1$, more protons are available for efficient interaction with
Alfven waves. Therefore, most of the turbulence energy goes to protons before
it reaches the proton gyroscale.
However, when $\beta_i\sim 1$, heating of protons is marginal and
most of the turbulence energy will cascade down further, crossing the proton gyroscale
(see Quataert 1999; Quataert \& Gruzinov 1999).

The Alfven waves will be converted into EMHD waves (whistlers) 
below the proton gyroradius scale.
When EMHD turbulence is anisotropic (i.e. $k_{\|}\ll k_{\perp}$), 
heating of protons by EMHD turbulence will be
marginal because ``protons sample a
rapidly varying electromagnetic field in the course of a
Larmor orbit (Quataert \& Gruzinov 1999)."
In this paper, we confirmed that anisotropy of EMHD turbulence ($k_{\|}\propto k_{\perp}^{1/3}$)
is strong. 
Therefore, in this case, the remaining energy (that is, the energy that has survived
the collisionless damping before
the proton gyroradius scale) cannot heat protons.
Instead, it will heat electrons when it reaches the electron gyroradius scale.
The bottom line is that the strong anisotropy of EMHD turbulence will make 
heating of protons by EMHD turbulence rather
difficult in collisionless astrophysical plasmas with $\beta_i \lesssim 1$
(see Quataert \& Gruzinov 1999).

\subsection{Other implications}

Other applications of the EMHD model include collisionless magnetic reconnection 
in laboratory and space plasmas (Bulanov, Pegoraro, \& Sakharov 1992;
Biskamp, Schwarz, \& Drake 1995; Avinash et al. 1998). Turbulence on microscale may be a source of anomalous resistivity, which can stabilize X-point reconnection. 
How important this type of reconnection is is a subject of debates. 
For instance, a model of magnetic reconnection in Lazarian \& Vishniac (1999) appeals to the magnetic field weak stochasticity, rather than microphysical plasma
effects to explain fast reconnection. Simulations by Kowal et al. (2009) successfully tested the Lazarian \& Vishniac (1999) model, which may mean that in many astrophysically important cases the reconnection is fast irrespectively of the plasma properties.
The anomalous effects, however, may be important for the initiation of the reconnection when the original level of turbulence in the system 
is low. In addition, in a partially ionized gas where the field wandering, which is the key element of Lazarian \& Vishniac (1999) model, is partially suppressed, the anomalous plasma effects can be important for reconnection (Lazarian, Vishniac \& Cho 2004). Incidentally, in the latter paper it is predicted that MHD turbulence is not killed by neutral friction in the partially ionized gas, but it gets resurrected at the scales at which neutrals and ions decouple. The resurrected cascade involves
only ions and electrons, not neutrals, and may proceed as electron MHD cascade below the proton gyroscale.

In addition, properties of EMHD turbulence are important for understanding of physics of neutron star crusts 
(Cumming et al. 2004; Harding \& Lai 2004) and acceleration of particles in Solar flares (Liu, Petrosian, \& Mason
2006).
While in most cases we view the EMHD cascade as the continuation of the MHD cascade below the proton gyroradius, for the crust of a neutral star the EMHD turbulence can be present on much larger scales. The main assumption of the EMHD is that one can ignore the motions of protons. This is definitely the case of the neutron star's solid crust.

\section{Summary}
We have found the following results.
\begin{enumerate}
\item Electron MHD (EMHD) and Electron Reduced MHD (ERMHD) show
      identical scaling relations (both spectra and anisotropies).
\item High resolution EMHD simulation confirms $k^{-7/3}$ spectrum
      obtained by earlier studies (Biskamp, Schwarz, \& Drake 1996;
      Biskamp et al. 1999; Ng et al. 2003; CL04). The spectrum of electric field
      is consistent with $k^{-1/3}$ spectrum obtained by
      earlier studies (see Schekochihin et al. 2009; Howes et al. 2008a;
      Dmitruk \& Matthaeus 2006).
\item Our detailed study of anisotropy using different techniques  
      supports $k_{\|}\propto k_{\perp}^{1/3}$ the EMHD scaling
      obtained by CL04. 
\item Decaying EMHD turbulence and driven EMHD turbulence show
      the same scaling.
\item When we use three or larger number of points to define 
      the structure functions, the scaling exponents of high-order structure functions follow
      a scaling similar to that of incompressible hydrodynamic turbulence.
\item Bispectrum of ERMHD turbulence, reflecting the coupling of different scales in the cascade,  looks very different from that of standard MHD one.
\item The probability distribution functions (PDFs) of the increment of the
      magnetic field strength in EMHD and MHD cases are well described by 
      the Tsallis distribution.
      The general trend of the PDFs is consistent with observations of
      the solar wind.
      
\end{enumerate}

\acknowledgements
We thank the anonymous referee for useful suggestions/comments. 
J.C.'s work was supported by the Korea Research Foundation grant
funded by the Korean Government (KRF-2006-331-C00136) and by KICOS through
the grant K20702020016-07E0200-01610 provided by MOST. 
A.L. acknowledges the support by the NSF grants ATM 0648699 and AST 0808118. Both authors
are supported by the NSF Center for Magnetic Self-Organization in Laboratory and Astrophysical 
Plasmas.


\begin{deluxetable}{llllccl} 
\tabletypesize{\footnotesize}
\tablecaption{Runs}  
\tablewidth{0pt}
\tablehead{
   \colhead{Run} & 
   \colhead{Resolution} & 
   \colhead{$B_0$}  &
   \colhead{$b$ at t=0}  &
   \colhead{$k_{\perp}$ at t=0\tablenotemark{\ddagger}}  &
   \colhead{$k_{\|}$ at t=0\tablenotemark{\ddagger}}  &
   \colhead{Comments} 
}
\startdata 
E512D        &  512$^3$         &    1 &  1.21 &   $2\le k < 5$ & $2\le k < 5$ & decaying \\
E256D        &  256$^3$         &    1 &  1.21 &   $2\le k < 5$ & $2\le k < 5$ & decaying \\
E256F        &  256$^3$         &    1 &  1.21 &   $k\sim 2.5$  & $k\sim 2.5$  & forced \\
E256D-EL & 768$\times$256$^2$   &  1 & 0.071 & $4.5/\sqrt{2}\le k_\perp \le 4.5\sqrt{2}$ & 
                                                      $1/3\le k_\|    \le 1$ & decaying \\
\hline
ER256D1-EL & 768$\times$256$^2$ &  1 & 0.071 & $4.5/\sqrt{2}\le k_\perp \le 4.5\sqrt{2}$ & 
                                                      $1/3\le k_\|    \le 1$ & decaying, $\sqrt{\alpha}=1$ \\
ER256D8-EL & 768$\times$256$^2$ &  1 & 0.071\tablenotemark{\dagger} & $4.5/\sqrt{2}\le k_\perp \le 4.5\sqrt{2}$ & 
                                                      $1/3\le k_\|    \le 1$ & decaying, $\sqrt{\alpha}=1/8$ \\
\hline
MHD512F        &  512$^3$      & 0.8 & 0     &   $k\sim 2.5$ & $k\sim 2.5$ & forced \\
MHD256F        &  256$^3$      &   1 & 0     &   $k\sim 2.5$ & $k\sim 2.5$ & forced \\
             \hline
\enddata
\tablenotetext{\dagger}{In this case, the value of $\tilde{b}$ at t=0 is listed.}
\tablenotetext{\dagger}{In cases of driven turbulence, the average driving-scale wavenumber is listed.}
\label{table_1} 
\end{deluxetable}


\appendix

\section{A. Derivation of ERMHD equations}
Schekochihin et al.~(2009) first derived ERMHD from kinetic RMHD equations.
They also gave derivation of ERMHD equations
{}from the generalized Ohm's law.
Therefore the starting point of ERMHD may be also the generalized Ohm's law.
Derivation of ERMHD is almost identical to the EMHD case.

Here we briefly summarize the derivation
of the ERMHD equation from the generalized Ohm's law.
Detailed derivation can be found in Appendix of Schekochihin et al. (2009).
The magnetic induction equation reads
\be
    \frac{\partial {\bf B}}{\partial t}=\nabla \times ({\bf v}_e \times {\bf B}),
   \label{eq:induct}
\ee
where
\be
   {\bf v}_e = {\bf v}_i -\frac{ {\bf j} }{en_e}
             = {\bf v}_i -\frac{c}{4\pi e n_e}\nabla \times {\bf B} \label{eq:ohms}
\ee
(see Appendix C of Schekochihin et al. 2009).
Here we ignored magnetic dissipation.
Note that the usual EMHD equations are also based on these equations.
Substituting Eq.~(\ref{eq:ohms}) into Eq.~(\ref{eq:induct}), we get
\bea
   \frac{\partial {\bf B}}{\partial t} & = &
           \nabla \times [ {\bf v}_i \times {\bf B} 
                -\frac{ c }{ 4\pi e n_{e} }(\nabla \times {\bf B})\times {\bf B} ],    
       \label{eq_a3} \\
    & = & -{\bf B}\nabla \cdot {\bf v}_i 
          -\frac{c}{4\pi e n_{e}}\nabla \times[(\nabla \times {\bf B})\times {\bf B}]
     -\frac{c}{4\pi e}\left(\nabla\frac{1}{n_e}\right)
      \times [(\nabla \times {\bf B})\times {\bf B}],
       \label{eq_a4}   \\
    & \approx & -{\bf B}\nabla \cdot {\bf v}_i 
          -\frac{c}{4\pi e n_{e}}\nabla \times[(\nabla \times {\bf B})\times {\bf B}],
       \label{eq_a5}   \\
   & = & -{\bf B}\nabla \cdot {\bf v}_i 
          -\frac{c}{4\pi e n_{e}}\nabla \times[{\bf B}\cdot \nabla {\bf B}],
\eea
which is the same as the EMHD equation except the ${\bf B}\nabla \cdot {\bf v}_i$ term on the right.
Here we assume ${\bf v}_i \approx 0$, but 
$\nabla \cdot {\bf v}_i=\nabla \cdot {\bf v}_e \ne 0$.
We ignore the last term of Eq.~(\ref{eq_a4})
because it is $\sim O(\epsilon^2)$. 
According to the RMHD ordering 
(see Section 2 of Schekochihin et al. 2009),
we have
\be
  \frac{\delta n_e}{n_e} \sim \frac{b}{B_0} \sim \frac{k_\|}{k_{\perp}} =\epsilon \ll 1,
\ee 
where $\delta n_e$ is the fluctuating density.

For $b_{\|}$ and ${\bf b}_{\perp}$, we have
\bea
   \frac{ \partial {\bf B}_{\perp} }{\partial t} =
      -\frac{c}{4\pi e n_{e}}\nabla_{\perp} \times[{\bf B}\cdot \nabla {\bf B}_{\|}], 
  \label{eq:perp3} \\
   \frac{ \partial {\bf B}_{\|} }{\partial t} =
      -{\bf B}_0 \nabla \cdot {\bf v}_i
      -\frac{c}{4\pi e n_{e}}\nabla_{\perp} \times[{\bf B}\cdot \nabla {\bf B}_{\perp}],
  \label{eq:par3}  
\eea
where we use $\nabla \times \approx \nabla_{\perp}\times$ 
and ${\bf B} \nabla \cdot {\bf v}_i
= {\bf B}_0 \nabla \cdot {\bf v}_i + O(\epsilon^2)$. Here $\epsilon \sim k_{\|}/k_{\perp}$.

{}From the continuity equation, we can rewrite the ${\bf B}_0 \nabla \cdot {\bf v}_i$ 
($={\bf B}_0 \nabla \cdot {\bf v}_e$) as
\bea
    {\bf B}_0 \nabla \cdot {\bf v}_i=
   -{\bf B}_0 \left( \frac{\partial}{\partial t}+{\bf v}_{\perp e}\cdot \nabla_{\perp}
              \right) \frac{\delta n_e}{n_{e}}  \\
   ={\bf B}_0 \left( \frac{\partial}{\partial t}+{\bf v}_{\perp e}\cdot \nabla_{\perp}
              \right)\frac{2}{\beta_i (1+Z/\tau)}\frac{b_{\|}}{B_0} \\
   = \frac{2 \hat{\bf z}}{\beta_i (1+Z/\tau)}\frac{\partial b_{\|}}{\partial t}
   +\frac{2}{\beta_i (1+Z/\tau)} \nabla_{\perp} \times
              ( {\bf v}_{\perp e} \times b_{\|}\hat{\bf z}) \\
   = \frac{2}{\beta_i (1+Z/\tau)}\frac{\partial {\bf b}_{\|}}{\partial t},
\eea
where we use ${\bf b}_{\|}=b_{\|}\hat{\bf z}$,
${\bf v}_{\perp e} \times b_{\|}\hat{\bf z}=O(\epsilon^2)\approx 0$, and 
$b_{\|}/B_0 = -(\beta_i/2)(1+Z/\tau)(\delta n_e/n_{e})$, which is
equivalent to the pressure balance.
Here, $Z=q_i/e$ ($e=|q_e|$) is the ion-to-electron charge ratio,
$\tau=T_i/T_e$ is the temperature ratio, 
$n_{e}$ is the mean electron number density,
$\delta n_e$ is the fluctuationing electron number density,
$\beta_i=8\pi n_i k_B T_i/B_0^2$ is the ion plasma beta.
Therefore, Eq.~(\ref{eq:par3}) becomes
\be
   \frac{ \partial {\bf B}_{\|} }{\partial t} =
      -\frac{\beta_i(1+Z/\tau)}{2+\beta_i(1+Z/\tau)}
      \frac{c}{4\pi e n_{e}}\nabla_{\perp} \times[{\bf B}\cdot \nabla {\bf B}_{\perp}].
\label{eq:par4}
\ee

In this paper, we use notations different from those in Schekochihin et al. (2009).
The notations in this paper and in Schekochihin et al. (2009) are related by
\bea
    {\bf b}          \leftrightarrow  \delta {\bf B},  \\
     b_{\|}          \leftrightarrow  \delta B_{\|}, \\
     {\bf b}_{\perp} \leftrightarrow   \delta {\bf B}_{\perp} \\
      n_{e} \leftrightarrow n_{0e}.
\eea

\section{B. Two-point second-order structure function and spectrum}
We can obtain 
the 2-point second-order structure function of a variable $A$
from $E_A(k)$:
\bea
   SF_2(r) & = & <|  A({\bf x}+{\bf r})-A({\bf x})|^2 > \\
     & = & \int d^3{\bf x} 
             \int d^3{\bf k} ~\hat{\bf u}(k) 
                 ~e^{i{\bf k}\cdot {\bf x}}(e^{i{\bf k}\cdot {\bf r}}-1)
             \int d^3{\bf k}^\prime ~\hat{\bf u}(k^\prime) 
                 ~e^{i{\bf k}^\prime \cdot {\bf x}}
                 (e^{i{\bf k}^\prime \cdot {\bf r}}-1) \\
      & = & \int d^3{\bf k} ~\hat{\bf u}(k) ~\hat{\bf u}^*(k)
            \left[ 2-e^{i{\bf k} \cdot {\bf r}}-e^{-i{\bf k} \cdot {\bf r}}
            \right] \\
      & = & 2\pi \int dk ~k^2 |\hat{\bf u}(k)|^2
                 \int d\theta ~\sin\theta 
            \left[ 2-e^{i{\bf k} {\bf r}\cos\theta}
                    -e^{-i{\bf k} {\bf r}\cos\theta}
            \right]  \\
      & = & 8\pi \int dk ~k^2 |\hat{\bf u}(k)|^2
                 \left[ 1-\frac{\sin kr}{kr} \right] \\
      & \propto & 8\pi r^{m-1} \int_{k_0r}^{\infty} d(kr)~(kr)^{-m}
                 \left[ 1-\frac{\sin kr}{kr} \right], \label{eq:B6}
\eea
where we assume 
\be
 k^2|\hat{\bf u}(k)|^2  \propto k^{-m}
\ee
for $k_0<k<\infty$.
In the limit of $kr \ll 1$, the integrand is proportional to 
$\sim (kr)^{2-m}$. 
Therefore, if $m < 3$, we can rewrite Eq.~(\ref{eq:B6}) as
\be
   SF_2(r) \approx 8\pi r^{m-1} \int_{0}^{\infty} d(kr)~(kr)^{-m}
                 \left[ 1-\frac{\sin kr}{kr} \right] \propto r^{m-1}.
\ee
If $m > 3$, however, the integral in Eq.~(\ref{eq:B6}) is roughly proportional to 
$(k_0r)(k_0r)^{2-m}=(k_0r)^{3-m}$ and we have 
\be
   SF_2(r) \propto r^{m-1}(k_0r)^{3-m} \propto r^2.
\ee
In summary, we have
\be
 SF_2(r) \propto \left\{ \begin{array}{ll} 
                    r^{m-1}    & \mbox{if $m < 3$} \\
                    r^2        & \mbox{if $m > 3$.}
                      \end{array}
              \right. \label{app:2ptsf2}
\ee
In case of Kolmogorov, $m=5/3<3$ and we have $SF_2(r) \propto r^{m-1}=r^{2/3}$.
In case of EMHD, $m$ for perpendicular direction ($m$=7/3)
is still smaller than $3$ and we have $SF_2(r) \propto r^{m-1}=r^{4/3}$.
However, that for parallel direction is expected to be larger than
$3$ if turbulence is anisotropic. Therefore, the 2-point second-order
structure function for parallel direction
is not suitable for revealing the true scaling exponent
 and 
we will have 
$SF_2(r) \propto r^{2}$ for parallel direction.

\section{C. Multi-point second-order structure function and spectrum}
Falcon et al.~(2007) and Lazarian \& Pogosyan (2008) used the 3-point
second-order structure function that can work with a steeper $E(k)$:
\bea
   SF_2(r) & = & 
       <|  A({\bf x}+{\bf r})-2A({\bf x})+A({\bf x}-{\bf r})|^2 > \\
     & = & 
             \int d^3{\bf x} 
             \int d^3{\bf k} ~\hat{\bf u}(k) 
                 ~e^{i{\bf k}\cdot {\bf x}}
                 (e^{i{\bf k}\cdot {\bf r}}-2+e^{-i{\bf k}\cdot {\bf r}})
             \int d^3{\bf k}^\prime ~\hat{\bf u}(k^\prime) 
                 ~e^{i{\bf k}^\prime \cdot {\bf x}}
    (e^{i{\bf k}^\prime \cdot {\bf r}}-2+e^{-i{\bf k}^\prime \cdot {\bf r}}) 
           \\
      & = & 
         \int d^3{\bf k} ~\hat{\bf u}(k) ~\hat{\bf u}^*(k)
            \left[ e^{i{\bf k} \cdot {\bf r}}-2+e^{-i{\bf k} \cdot {\bf r}}
            \right] 
            \left[ e^{i{\bf k} \cdot {\bf r}}-2+e^{-i{\bf k} \cdot {\bf r}}
            \right]^*
           \\
      & = & 2\pi \int dk ~k^2 |\hat{\bf u}(k)|^2
                 \int d\theta ~\sin\theta 
            \left[ e^{ 2i{\bf k} {\bf r}\cos\theta}
                  +e^{-2i{\bf k} {\bf r}\cos\theta}
                  +6
                  -4e^{ i{\bf k} {\bf r}\cos\theta}
                  -4e^{-i{\bf k} {\bf r}\cos\theta}
            \right]  \\
      & = & 8\pi \int dk ~k^2 |\hat{\bf u}(k)|^2
             \left[ \frac{\sin 2kr}{2kr} +3-4\frac{\sin kr}{kr} \right] 
               \\
      & \propto & 8\pi r^{m-1} \int_{k_0r}^{\infty} d(kr)~(kr)^{-m}
              \left[ \left(4-4\frac{\sin kr}{kr}\right)
                    -\left(1- \frac{\sin 2kr}{2kr}\right)  \right],
\eea
where we assume 
 $k^2|\hat{\bf u}(k)|^2  \propto k^{-m}$
for $k_0<k<\infty$.
In the limit of $kr \ll 1$, the integrand is proportional to 
$\sim (kr)^{4-m}$. 
Therefore, the integral is roughly a constant (i.e.~independent of $r$), if $m < 5$.
If $m > 5$, however, the integral is roughly proportional to 
$(k_0r)(k_0r)^{4-m}=(k_0r)^{5-m}$.
Therefore, we have
\be
 SF_2(r) \propto \left\{ \begin{array}{ll} 
                    r^{m-1}    & \mbox{if $m < 5$} \\
                    r^{m-1}(k_0r)^{5-m} \propto r^4  & \mbox{if $m > 5$.}
                      \end{array}
              \right. \label{app:3ptsf2}
\ee

Similarly, we can construct
a 4-point 
second-order structure function:
\bea
   SF_2(r) & = & 
       <|  A({\bf x}+3{\bf r})-3A({\bf x}+{\bf r})
                              +3A({\bf x}-{\bf r})
          -A({\bf x}-3{\bf r}) |^2 >, \mbox{~~~or} \\
    SF_2(r)  & = & 
       <|  A({\bf x}+\frac{3}{2}{\bf r})-3A({\bf x}+\frac{1}{2}{\bf r})
                              +3A({\bf x}-\frac{1}{2}{\bf r})
          -A({\bf x}-\frac{3}{2}{\bf r}) |^2 >,
\eea
which scales as
\be
 SF_2(r) \propto \left\{ \begin{array}{ll} 
                    r^{m-1}    & \mbox{if $m < 7$} \\
                    r^{m-1}(k_0r)^{7-m} \propto r^6  & \mbox{if $m > 7$.}
                      \end{array}
              \right. \label{app:4ptsf2}
\ee
We can also construct
a 5-point 
second-order structure function:
\be
   SF_2(r) = 
       <|  A({\bf x}+2{\bf r})-4A({\bf x}+{\bf r})+6A({\bf x})
                              -4A({\bf x}-{\bf r})
          -A({\bf x}-2{\bf r}) |^2 >,
\ee
which scales as
\be
 SF_2(r) \propto \left\{ \begin{array}{ll} 
                    r^{m-1}    & \mbox{if $m < 9$} \\
                    r^{m-1}(k_0r)^{9-m} \propto r^8  & \mbox{if $m > 9$.}
                      \end{array}
              \right. \label{app:5ptsf2}
\ee

\end{document}